\documentclass[lettersize,journal]{IEEEtran}
\usepackage{amsmath,amsfonts}
\usepackage{algorithmic}
\usepackage{algorithm}
\usepackage{array}
\usepackage[caption=false,font=normalsize,labelfont=sf,textfont=sf]{subfig}
\usepackage{textcomp}
\usepackage{stfloats}
\usepackage{url}
\usepackage{verbatim}
\usepackage{cite}
\usepackage{hyperref}
\usepackage{graphicx}
\usepackage{booktabs}
\usepackage{multirow}
\usepackage{makecell}

\usepackage[T1]{fontenc}
\usepackage[table]{xcolor} 
\definecolor{shaded}{HTML}{E6E2DE}

\begin{document}

\title{Exploring spatial configuration and \\group familiarity when co-watching \\360-degree stories in social virtual reality}

\author{Portia Wang, Monique Santoso, Yujie Tao, and Jeremy Bailenson
        % <-this % stops a space
\thanks{}
\thanks{Portia Wang, Monique Santoso, Yujie Tao, and Jeremy Bailenson are with Stanford University, Stanford, CA, 94305, USA (email:portiaw@stanford.edu; msant@stanford.edu; yjtao@stanford.edu; bailenson@stanford.edu).}}% <-this % stops a space}

\markboth{}
{Wang \MakeLowercase{\textit{et al.}}}

%\IEEEpubid{0000--0000/00\$00.00~\copyright~2026 IEEE}
\maketitle

\begin{abstract}
While traditional cinematic viewing is often characterized as a social experience, engaging with immersive stories is predominantly researched as an individual activity. This mismatch becomes increasingly problematic in social virtual reality (VR), where features designed for typical social activities (e.g., virtual locomotion, spatial and social context) can become meaningful levers for shaping the user experience. To address our limited understanding of these effects, we conducted three exploratory experiments (N$_1$=109, N$_2$=116, N$_3$=88) examining how spatial boundary type, size of allocated personal space, and group familiarity influence co-watching 360-degree stories in social VR. Results shows that small communal spatial boundaries heightened social presence towards others compared to individual boundaries, but also increased mental demand. Reducing the size of the personal spatial boundary increased social presence towards others but also increased disorientation and frustration. Finally, participants with higher group familiarity reported stronger attentional focus on the story, as well as higher spatial presence and social presence towards story characters, but also reported greater physical and perceptual demand. We discuss systematic tradeoffs between engagement and cognitive effort, and suggest practical guidelines for designing shared immersive narrative experiences.
\end{abstract}


\section{Introduction}
Engaging with cinematic stories is often a social act. We go to theaters not only to watch stories unfold, but to laugh when others laugh and flinch when others do. This social dimension extends beyond theaters, as friends watch movies together in private spaces or through online watch parties, and strangers co-experience content through live or asynchronous comment features \cite{Chen2017, Weisz2007}. Watching a story alone can be powerful, but doing so together can change what it means to experience it. Indeed, the popularity of huge, multisensory immersive domes such as the Sphere in Las Vegas, which entertains thousands of people at once, demonstrates how closely tied immersion and social interaction are \cite{Schmitt2026}. 

Virtual reality (VR) shapes the social and spatial dynamics of groups who join together virtually to view cinematic stories \cite{Herscher2019, Layng2019}. For one, the technology immerses users in the same virtual environment when watching cinematic stories, allowing them to embody avatars and to interact with others in ways that are analogous to real-world interactions \cite{Maloney2020, Smith2018}. Moreover, virtual environments can project 360-degree videos and facilitate interactive experiences that surround users with story elements from all angles \cite{Godde2018}. This differs from previous approaches that simulate theater-like experiences by projecting 2D films onto virtual screens \cite{Moss2024}. In co-watching 360-degree stories through immersive platforms, social others can occupy a more perceptually salient role, though how we should instantiate these virtual experiences remains an open question.

Consider the following hypothetical scenario. Alone in their living room, a user puts on a headset to join friends in a networked VR environment to watch a 360-degree story together. At first, the user's avatar clusters near their friends, eager to connect with those in the environment. Yet, as the story unfolds, the avatars of nearby users begin to obstruct parts of the spherical scene, and the simultaneous motion of story characters and user avatars becomes visually overwhelming. The user translates away to regain a clearer view using their controller, but in doing so, accidentally clashes into a group of avatars standing behind them. The virtual collision triggers a small domino effect: one avatar shies sideways, another pivots to avoid them, and the group collapses into a scramble of repositioning. For a brief moment, everyone’s attention shifts away from the story to managing the social disruption. Eventually, the user manages to teleport themselves away from the perceptual cluster of moving avatars: the avatars of their friends become smaller in the field of view, until they are no longer visually discernible. When the story ends, they realize that they knew little of how their friends reacted and felt. What began as an engaging social activity now feels solitary. 

As illustrated through this scenario, co-watching in immersive environments is inherently complex, requiring ongoing negotiation of space and attention. The themes it surfaces are reflected in empirical VR work outside the co-watching context, where researchers have identified virtual space negotiation \cite{Williamson2021, Welsch2021, Williamson2022}, avatar collision \cite{akselrad2023, wang2025tvcg}, and visual occlusion and clutter \cite{hu2025, yu2019, wolf2022} as key considerations for designing meaningful immersive experiences that are not cognitively overloading. These dynamics are also unlikely to be uniform across social groups: users stand closer to those they are more familiar with in virtual spaces \cite{Kim2024}, so a group of strangers and a group of friends may arrive at drastically different spatial and attentional dynamics while watching the same 360-degree story. That said, how space and attention are negotiated during co-watching, and how social relationships shape this negotiation, remains largely unexplored.

This work investigates how these factors influence immersive co-watching of 360-degree stories by studying how different instantiations of the experience influence psychological outcomes. We conducted three experiments (N$_1$=109, N$_2$=116, N$_3$=88) where groups of individuals gathered in social VR to co-watch immersive cinematic stories. The first two experiments examined the spatial configuration of co-watching. Experiment 1 varied \textit{virtual spatial boundary}, which refers to how the virtual environment assigns space to its co-watching users, and drew on real-world use cases that differ in how much space is available and in whether that space is communal (i.e., users share a single region) or personal (i.e., each user is allocated a separate region that others cannot enter). Experiment 2 then narrowed to the personal space configuration, holding the spatial boundary type constant and varying the size of each user's allocated \textit{personal space}. The third experiment turned to the social dimension of co-watching, exploring how group familiarity is related to psychological and behavioral outcomes. The design of a co-watching space does not determine who occupies it, so we examine the spatial and social dimensions as complementary and equally important aspects of the user experience. Accordingly, our work is guided by two research questions:

\begin{itemize}
   \item \textbf{Research Question 1 (RQ1)}: How do spatial configurations influence one’s subjective experience and behavioral patterns during immersive co-watching of 360-degree movies? Specifically, what are the effects of varying the type of virtual spatial boundary (\textbf{RQ1a}) and the size of allocated personal space (\textbf{RQ1b})?
   \item \textbf{Research Question 2 (RQ2)}: How does one’s familiarity with their co-watchers influence their subjective experience and behavioral patterns during immersive co-watching of 360-degree movies?
\end{itemize}

This work is exploratory as this design space is largely underexamined, with little empirical evidence from which to derive directional hypotheses. Our large samples drawn from a university course served to estimate effects rather than to test confirmatory, pre-registered predictions. By identifying systematic tradeoffs between social and narrative engagement alongside effort and workload, we contribute to the understanding of immersive co-watching experiences and discuss practical design implications for doing so. 

\section{Background}
\subsection{Theoretical Perspectives of Watching Immersive Stories}

Researchers have examined the psychology of watching cinematic stories by analyzing how viewers allocate attention across content of varying levels of immersion. Early research revealed that people tend to look at similar parts of 2D movie clips, even when doing so individually rather than in groups \cite{Goldstein2007}. The same is true for individuals watching immersive 360-degree stories, which Bender \cite{Bender2019} referred to as attentional synchrony. Despite 360-degree stories affording users the ability to actively change their viewing perspective unlike in 2D movies, attention often converges across viewers, which researchers have attributed to their shared strategies of using centers of interest and salient perceptual cues \cite{Bender2019, Goldstein2007}. 

Introducing co-watching complicates these dynamics. For one, co-watchers can function as competing or reinforcing attentional cues, diverting attention away from the narrative or directing it through gaze following and shared focus. One helpful theoretical framework is the model of narrative engagement \cite{Busselle2008Model}, which conceptualizes narrative engagement as the construction and continual updating of mental models for understanding story narratives. The theory posits that built mental models require sustained cognitive effort and attentional focus, often accompanied by a reduced awareness of oneself and the immediate surroundings. In the context of immersive co-watching (i.e., co-watching through immersive technology), it is likely that the cognitive effort required to engage with immersive stories adds to the workload for being socially present with co-watchers. Conversely, co-watchers could also effectively shape attention by guiding users to look at visually salient areas, in turn lowering cognitive demand. 

Co-watching also redefines viewing immersive stories as a social experience, where one’s subjective sense of being there with others (i.e., social presence) becomes an important factor. While Oh et al. \cite{Oh2018} suggest that social presence can be enhanced by using behaviorally realistic virtual avatars, it is unclear how these social cues interplay with media consumption. Unlike prior research that foregrounds social interaction or collaboration as the primary task \cite{Maloney2020, Moustafa2018}, co-watching presents a tension between the primary task of narrative engagement and the secondary need to engage in social negotiation and maintenance. Being in a virtual environment with other users can also alter how individuals perceive virtual characters in immersive stories. For example, Daher et al. \cite{Daher2017} found that participants who observed realistic interactions between a confederate and a virtual human reported greater social presence and attentional allocation towards the virtual human compared to those who did not receive this priming. It is therefore possible that one’s perception of how co-watchers interacted and reacted to story characters can influence their perception of the characters and narrative story as a whole. 

Taken together, past research provides a helpful starting point in understanding the social dynamics of immersive co-watching, though the lack of empirical work on this topic leaves practitioners with limited guidance on how to effectively design co-watching experiences. Our work bridges this gap through three experiments that examine the psychological and behavioral effects of co-watching 360-degree stories in VR. Specifically, we examined co-watching experiences through two types of context: spatial and social.

\subsection{Spatial Design of Co-watching Experiences}
Spaces shape how people interact with them and with each other. In non-VR settings, spatial arrangements such as seating layouts can influence engagement and group behavior \cite{Wannarka2008}. Similar effects were observed in digital environments, where the spatial configuration of virtual spaces communicates social norms and expectations for interaction \cite{Stromer2009}. Spatial layouts not only afford different forms of participation but also convey cues about how a space is to be occupied and used \cite{Han2025cuddle, Osborne2023, Han2024chi}. 

How individuals choose to move and linger in spaces can also signal underlying mechanisms governing social interactions. Scholars have prominently considered interpersonal distance through the proxemics theory \cite{Hall1966}, which posits that interpersonal distance is governed by zones of distance, ranging from a close intimate zone to a bigger public zone, and reflects psychological states and social relationships \cite{Kim2024, Nolte2025}. Empirically, closer proximity between individuals in VR positively correlates with social presence \cite{Oh2018}, though avatars standing exceedingly close to an individual may elicit a sense of threat \cite{Wang2024}. Restricting virtual movement can further change how individuals behave and collaborate, as spatially constrained users have been found to produce less practical designs and fewer iterations \cite{Han2025}. Researchers have also referenced norms in social interactions when designing spatial configurations for social gatherings; Kiuchi et al. \cite{Kiuchi2025} and Wang et al. \cite{Wang2024} adopted the circular formations seen in social interactions to design spatial configurations for augmented and virtual reality meetings, while Qiu et al. \cite{Qiu2023} and Wong et al. \cite{Wong2024} did so for users meeting in mixed reality.

Co-watching 360-degree stories introduces novel dynamics. For one, while researchers have considered the placement of 2D screens and whiteboards for collaboration in VR \cite{Moss2024, Osborne2023, Wang2024mapping}, these elements are largely fixed in virtual spaces and do not require continuous, wide-range head turning. Attending 360-degree content is inherently different, as users are often required to look around to follow the storyline \cite{Jun2020, Godde2018, Kim2022}. Moreover, movement in the virtual space also does not alter the viewpoint of the story, as the story is typically rendered through skyboxes. Instead, virtual movement changes only how co-watchers are spatially arranged relative to the story and to one another. This gives users opportunities to avoid others from occluding their view of story elements and align their movement to the camera movements so that other users appear more diegetic to the story world. That said, the presence of other users may also introduce non-diegetic perceptual conflicts; when avatars occupy spatial positions that are inconsistent with the story world (e.g., standing too far), they undermine the internal consistency of the immersive experiences \cite{Gottsacker2021}. To better understand these new dynamics, our first two experiments manipulate the spatial boundary type and the size of allocated personal spaces (\textbf{RQ1}) and identify how spatial context shapes co-watching of 360-degree stories.

\subsection{Social Context Shaping Immersive Experiences}\label{sec: related work social context}
Viewing experiences can also vary across social contexts. In face-to-face scenarios, consumption of cinematic stories ranges from private viewings with friends to public screenings among strangers. This social dimension is mirrored in online contexts, where watch parties and social VR allow users to co-experience content with people \cite{Luo2020, Sun2017}. Prior works on co-watching in non-immersive online settings found that viewing comments left by other viewers and experiencing content socially evoked stronger emotional expressions \cite{Luo2020}, yielded more synchronized brain activity \cite{De2024}, and built social connection \cite{Chen2017}. By contrast, there is less research directly informing how prior relationships shape immersive co-watching. 

One relevant line of research examines how group familiarity shapes social dynamics and collaboration, showing that social closeness is positively associated with creativity \cite{Oztop2018}, satisfaction in collaboration \cite{Feng2022}, and willingness to share \cite{Wiese2011}. High-familiarity groups exhibit greater emotional contagion \cite{Du2014} and require less active regulation of social and task processes \cite{Janssen2009}. Because familiar group members share a social history, more familiar groups can establish group norms more easily and devote less time to regulating and monitoring task-related activities \cite{Janssen2009}. In the context of co-watching, where story engagement constitutes the primary task, this reduction in social regulation effort for high familiarity groups may permit a greater allocation of attentional resources toward the task.

A competing psychological mechanism suggests that familiarity may distract individuals from immersive stories. Known as familiarity bias \cite{Fernandez2025}, the heuristic posits that individuals tend to interact more frequently and rely more heavily on group members with whom they have existing relationships. This is evident in VR, as friends tended to engage in more turn-taking and expressed higher co-presence compared to strangers \cite{Tutuncu2025}. Users familiar with their co-watchers might therefore prioritize attention towards social interaction over the story.

These conflicting possibilities reveal a significant knowledge gap in how practitioners should design for immersive co-watching across social contexts. For example, an intimate virtual spatial configuration designed for friends might increase the cognitive burden and discomfort for a group of strangers, who may in turn feel a greater need to monitor social dynamics rather than to focus on the narrative. As we envision a future where social VR platforms can support both communal viewing among close peers and strangers, it is vital to understand how varying degrees of interpersonal familiarity shape psychological and behavioral measures such as presence, workload, and interpersonal distance. Our second research question and final experiment address this gap.

\begin{figure*}[t]
 \centering 
 \includegraphics[width=0.9\textwidth,keepaspectratio]{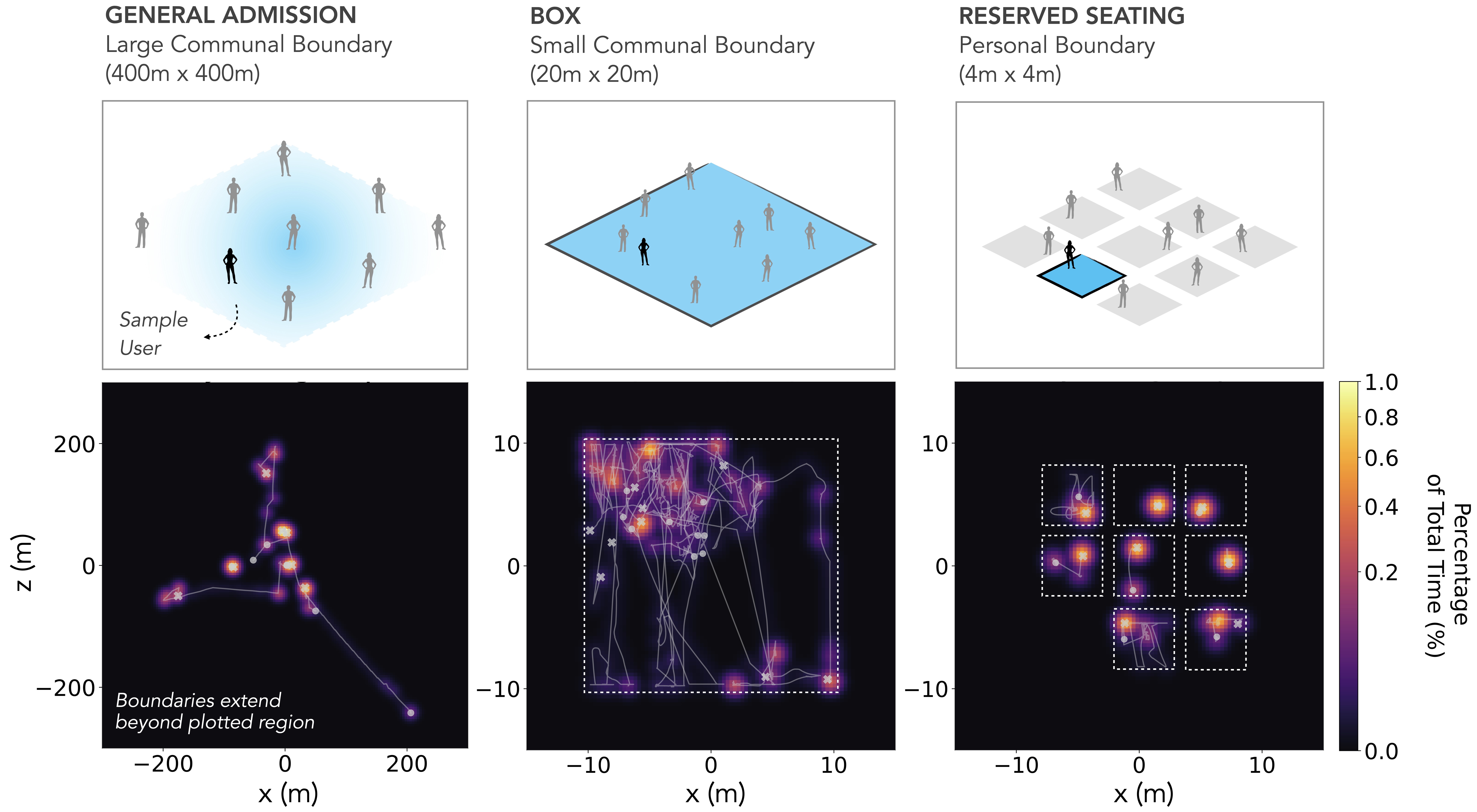}
 \caption{Examples of Spatial Configurations and Behavioral Patterns. Top panels show schematics of virtual spaces examined in Experiment 1, with black outlines and blue overlays denoting the spatial constraints of the sample user. Bottom panels visualize top-down movement patterns from the experiment. White dotted outlines indicate virtual boundaries and heatmap denote group-level spatial dwelling patterns. Virtual boundaries are slightly expanded to increase the visibility of movement patterns. Gray paths denote individual movement patterns, with dots and ``x''s indicating users' start and end positions, respectively.}
 \label{fig: movement visualization}
\end{figure*}

\subsection{Motivations for Prototyping Spatial Configurations}
\label{section: motivation for spatial configurations}
Immersive co-watching introduces complexities to how practitioners may instantiate these experiences. How should users be placed, where should they be allowed to go, and does it matter who they co-watch with? While prior works suggest that examining self-report group familiarity addresses the final question (\textbf{RQ2}), how we should vary spatial configurations remains abstract (\textbf{RQ1}). In this section, we explicate the rationale behind the Experiments 1 and 2 conditions.

Our earlier example of a user joining an immersive co-watching session with friends highlighted a few spatial considerations. The first is spatial boundaries (\textbf{RQ1a}). To study this, we grounded Experiment 1 conditions by referencing current practices of immersive co-watching. Default spaces used for spatial co-watching in immersive contexts usually take one of two forms. In social VR platforms such as ENGAGE XR \cite{ENGAGE}, users are often placed in largely unconstrained 360-degree video rooms with high movement autonomy (i.e., large communal boundaries). This lack of spatial constraints, however, can lead to users scattered throughout virtual spaces, unable to effortlessly engage in meaningful social interactions. When they do manage to cluster around others, actively coordinating proxemics while watching immersive stories can introduce unintended violations of social norms (e.g., colliding avatars). 

Another common spatial configuration mirrors those seen in physical theaters \cite{Moss2024}, where users are spatially bounded to specific locations in virtual space (i.e., personal boundary). Visual indicators of the spatial constraints are often instantiated through virtual seats \cite{Moss2024, Layng2019}, with user movements largely constrained to head rotations. This latter form of spatial configuration often mandates high compatibility between the narrative content and seating arrangements, for example, having a virtual screen paired with tiered seating to show 2D movies \cite{Moss2024} or instantiating customized seating arrangements to tailor to the story's narrative arc \cite{Layng2019, Herscher2019}. Occlusion becomes a more salient problem in this configuration, with users now unable to freely move in the virtual space to gain a clearer vantage point of story content. From the two spatial configuration types of large communal boundary and personal boundary, we identified an in-between, where the virtual environment is communal, yet reasonably constrained. Taken together, these three types of spatial boundaries (i.e., large communal, small communal, personal) make up the experimental conditions in Experiment 1. Of these, small communal and personal boundaries are already found in real-world immersive venues, such as the Las Vegas Sphere and Cosm Los Angeles. Figure \ref{fig: movement visualization} offers an overview of how the three spatial boundaries corresponded to distinct behavioral patterns in our experiment.

\begin{figure*}[t]
 \centering 
 \includegraphics[width=0.9\textwidth,keepaspectratio]{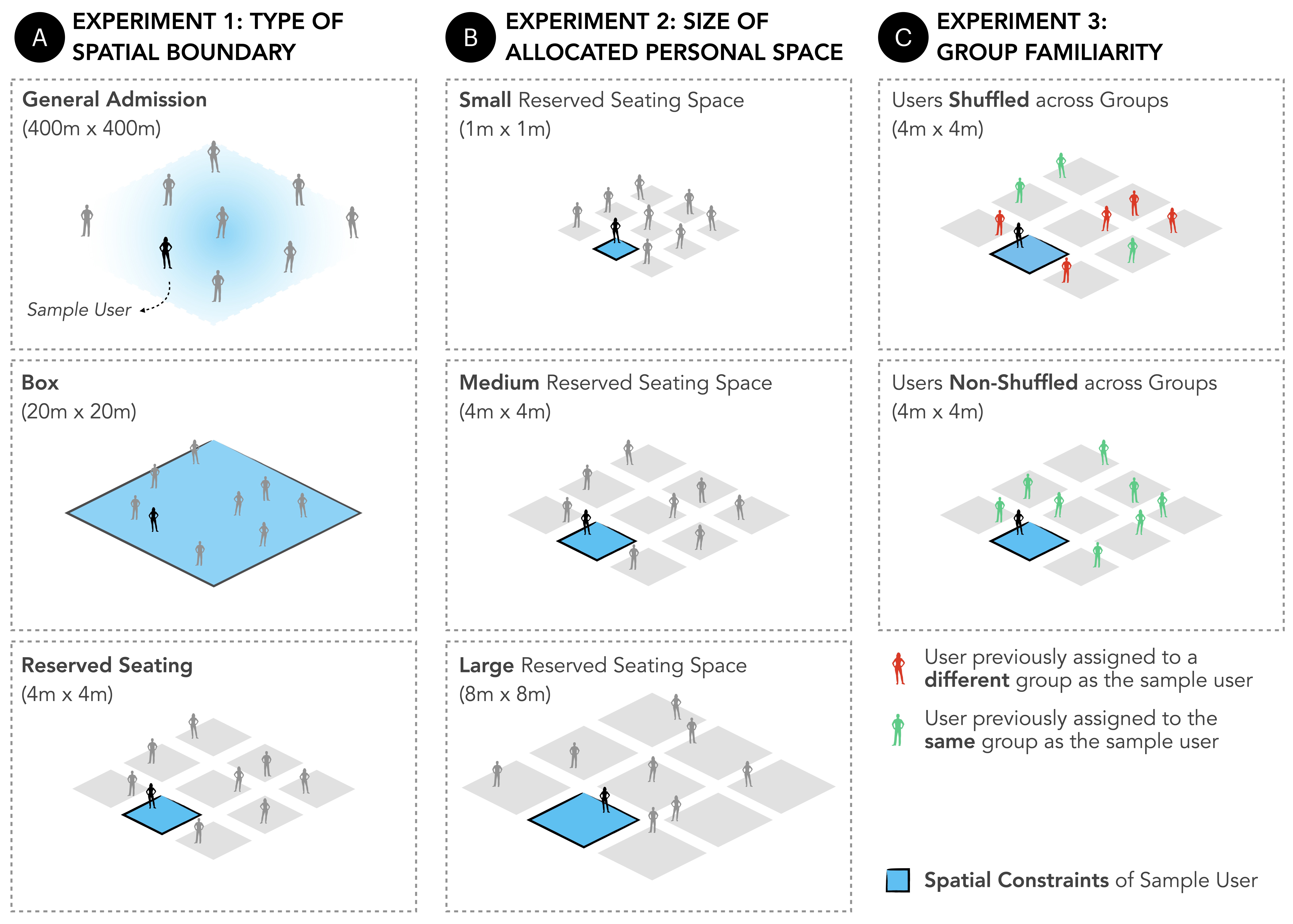}
 \caption{Schematics of the Virtual Spaces. Sample user is allowed to move within the outlined blue spatial constraints using either teleportation or smooth translation. In practice, to accommodate different-sized groups, the reserved seating configurations allow for up to 16 participants, using a four-by-four grid.}
 \label{fig: conditions}
\end{figure*}

Experiment 2 takes a deep dive into the dimension of personal space, posing the question of how big the personal space should be (\textbf{RQ1b}). This inquiry is driven by three key practical and theoretical motivations. First, personal spaces and boundaries offer a highly scalable design solution: while the ideal size of a visual space with communal boundaries will likely depend heavily on the size of a social group (e.g., bigger groups warranting larger spaces), a standardized personal boundary provides a more consistent ``building block'' that can be deployed across social contexts. Furthermore, by referencing the fixed seating arrangements of physical theaters, we tap into a spatial configuration that is not only intuitive to navigate for novice users but also deeply rooted in prior research on mobility \cite{Han2025}, personal space \cite{Pohl2019}, and more broadly proxemics \cite{Williamson2021, Williamson2022,Welsch2021}. Finally, leveraging instances of personal boundaries reduces within-group variability in behavioral patterns such as interpersonal distance for those assigned to the same condition. In more loosely-constrained spatial configurations (e.g., large communal boundary), individual movement and spatial negotiations can vary considerably across group members and groups assigned to the same condition such that it becomes difficult to compare across spatial designs. Partitioning the virtual space using individual boundaries (see third column of Figure \ref{fig: movement visualization} for an example) allows us to more directly investigate how behavioral patterns shaped by spatial characteristics (i.e., size of personal space, interpersonal distance) are related to psychological outcomes such as presence, attention, and cognitive effort.

\section{Methods}
We conducted three experiments examining the effects of spatial boundary type, size of personal space, and group familiarity on co-watching 360-degree stories in VR (Figure \ref{fig: conditions}). Using participants drawn from a university course, the experiments each followed a between-subjects design, manipulating their examined variables at two or three levels. The study protocol is pre-registered using \href{https://osf.io/zbr2d/overview?view_only=5b40d2032a1b4df39df63b1eeb1cf17c}{OSF}. We next detail the methods and study protocol shared across the three experiments, before elaborating on experiment-specific demographics, conditions, and results in Sections \ref{sec: exp 1}--\ref{sec: exp 3}.

\subsection{Participants}\label{sec: participants}
We recruited participants through a university course on VR. The study protocol was approved by the university’s Institutional Review Board (IRB) and a second university ethics organization. The research teams were blind to the consenting status of students until the end of the course to ensure that all students received the same pedagogical experiences regardless of consenting status. 
%Aligned with the IRB protocol, all students were required to complete the self-report questionnaires for pedagogical purposes, and responses from non-consenting students were excluded from subsequent data analysis. A third-party coordinator not part of the teaching or research team administered logistical procedure that warranted access to the consenting status, such as in scheduling timeslots to maximize for the number of students in all-consenting groups. 
The three experiments were examined separately, with sample sizes of 109, 116, and 88, respectively. We detail their demographics in Sections \ref{sec: exp 1}--\ref{sec: exp 3}.

Participants were then assigned to nine time slots. We divided participants from the same time slot into two groups, which allowed us to manipulate group familiarity in Experiment 3. \footnote{Given spatial constraints, two out of the nine time slots were kept as a single group with roughly half the participants assigned to other timeslots.} Groups had between five and ten participants (M$_1$ = 8.31, SD$_1$ = 1.35; M$_2$ = 8.50, SD$_2$ = 1.37; M$_3$ = 7.73, SD$_3$ = 1.28). Throughout the study, groups gathered in the same physical space and completed all onboarding and co-watching activities together. Participants who were unable to show up for a particular experiment were not permitted to swap into other groups, but were allowed to participate in later experiments. 

\subsection{Study Setup and Equipment}
Participants each used a Meta Quest 2 headset and two hand-held controllers to partake in the co-watching experiences that were hosted on the social VR platform ENGAGE XR. 360-degree stories were loaded using the platform’s video player and rendered as skyboxes. The platform enabled virtual movement through physical movement (e.g., head turning), as well as smooth translation, teleportation, and 30-degree yaw rotations using controllers. Notably, both physical and virtual head rotations change the story viewpoint, while translation does not. We leveraged the platform’s recording tool to record the VR sessions, and its 3D modelling tools for creating spatial configurations. Using Python scripts that format the positions, rotations, and scales of 3D objects into the IFXSnapshot structure, we were able to import custom spatial elements such as visible ground markers for users to stand on or invisible colliders to form spatial boundaries.

\begin{table*}[t]
\centering
\caption{The Five Sampled Content. The first three pieces are used in Experiments 1 and 2, and the final two are used in Experiment 3.}
\label{tab:content}
\vspace{0em}

\renewcommand{\arraystretch}{1.02} % Increased slightly for better breathing room

\begin{tabular}{p{2.5cm} p{1.2cm} p{3.5cm} p{3.5cm}} % Changed 'm' to 'p' for top-alignment
\toprule
\textbf{Content} & \textbf{Duration} & \textbf{Description} & \textbf{360-Degree Screenshot} \\
\midrule

ARIPI & 8:09 & 
An astronaut finds himself trapped in a failing spacecraft, returning to his childhood imagination as the spacecraft falls to Earth. & 
\vspace{-1em} \includegraphics[width=\linewidth]{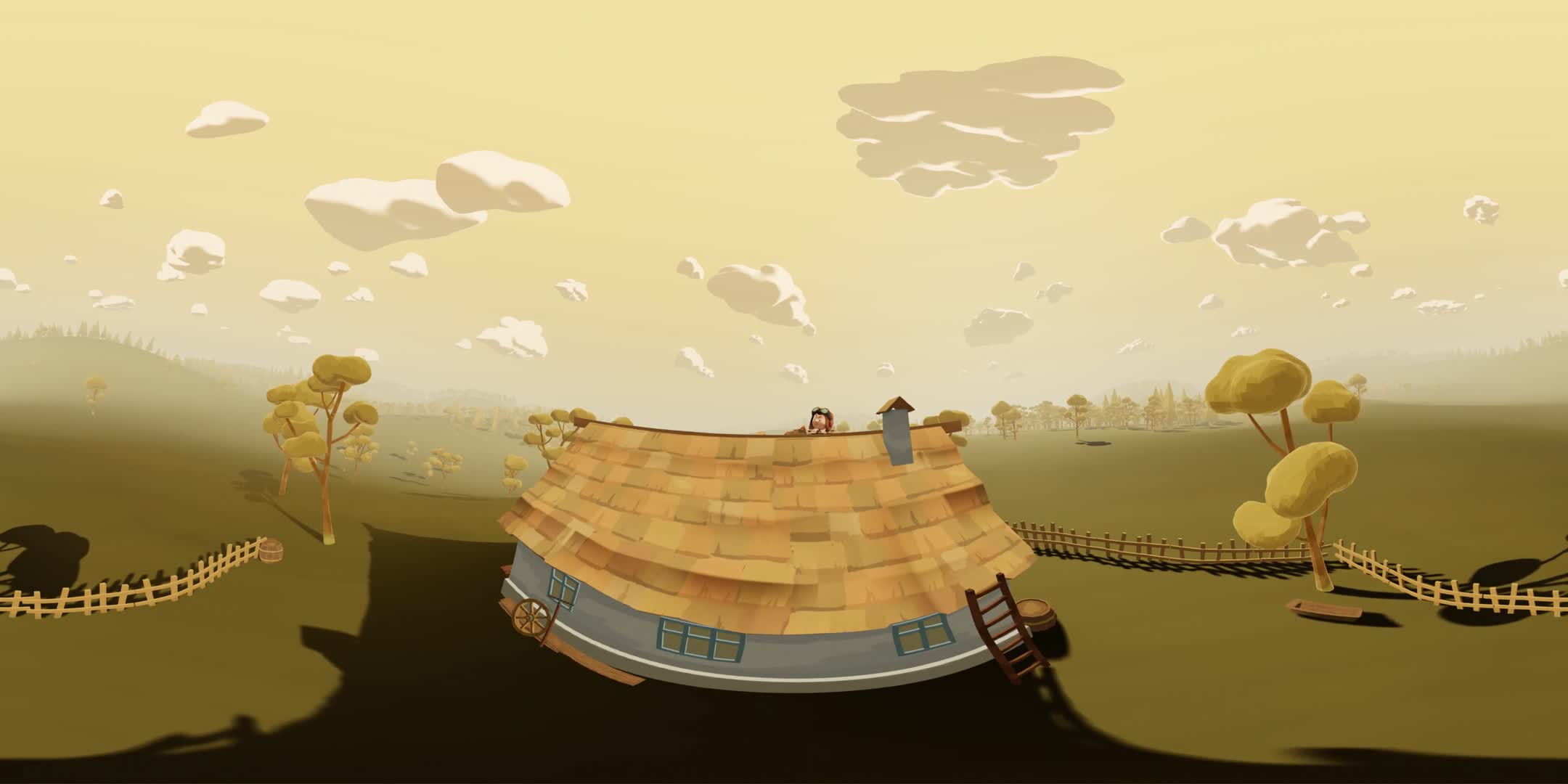} \vspace{-1em} \\

Help & 4:54 & 
A meteor shower triggers chaos in the city as an alien chases a woman through the subway and into the streets. & 
\vspace{-1em} \includegraphics[width=\linewidth]{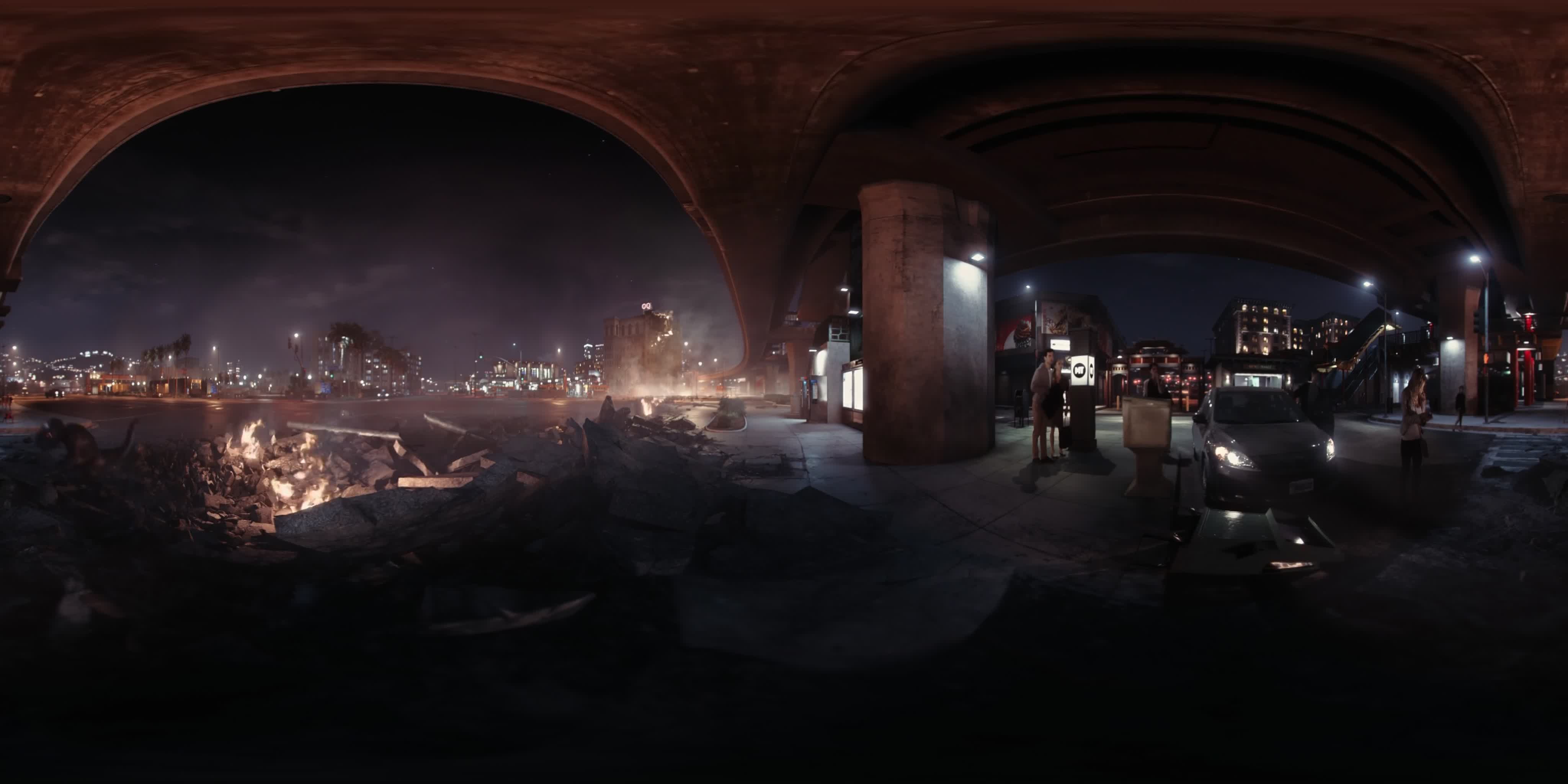} \vspace{-1em} \\

Rain or Shine & 5:29 & 
A young girl receives a package from a deliveryman and discovers the unexpected power of her new sunglasses. & 
\vspace{-1em} \includegraphics[width=\linewidth]{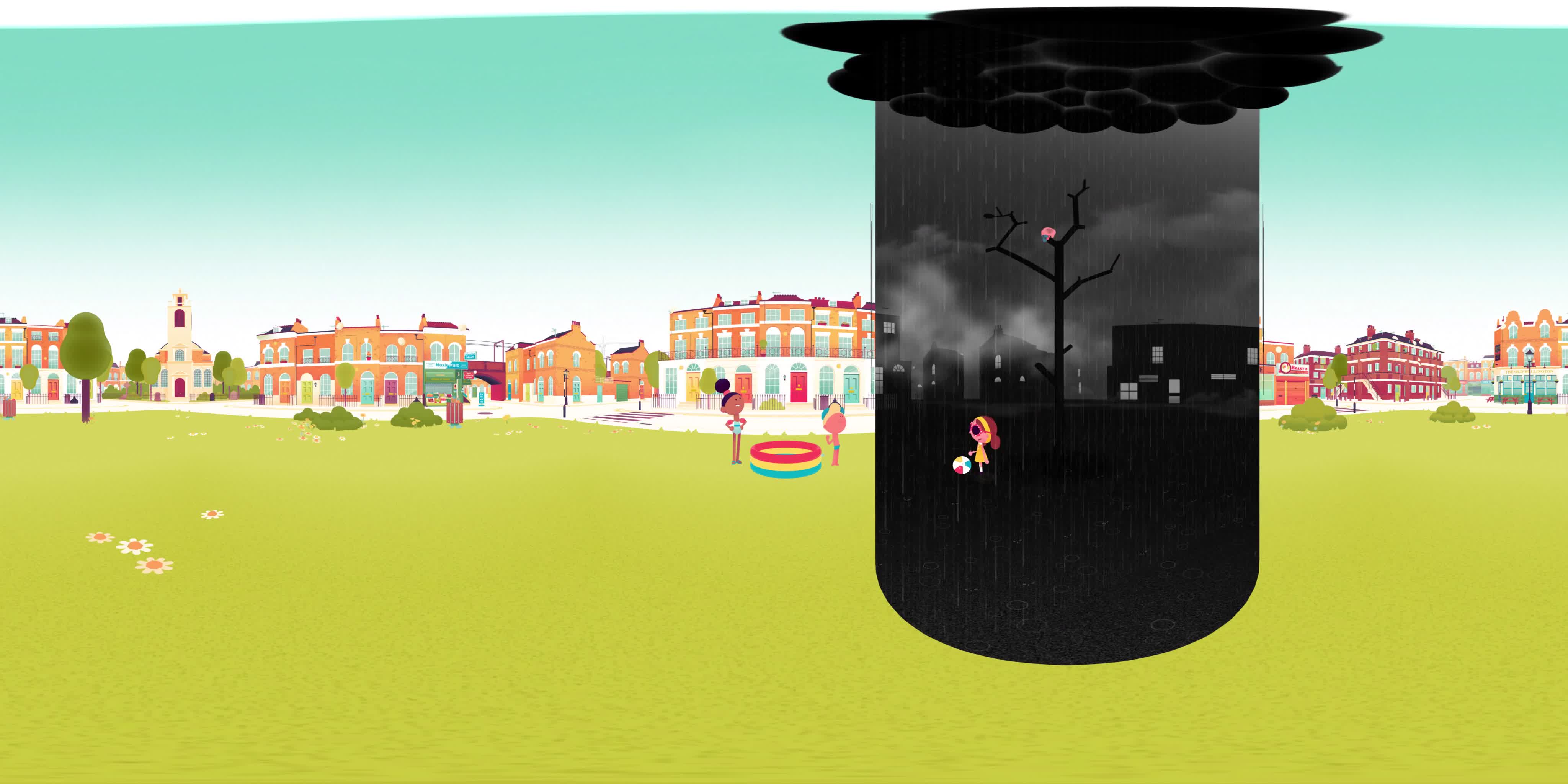} \vspace{-1em} \\

Journey into the History of Cinema & 9:32 & 
A ride through the history of cinema, with animations of classic films adapted and recreated in virtual reality. & 
\vspace{-1em} \includegraphics[width=\linewidth]{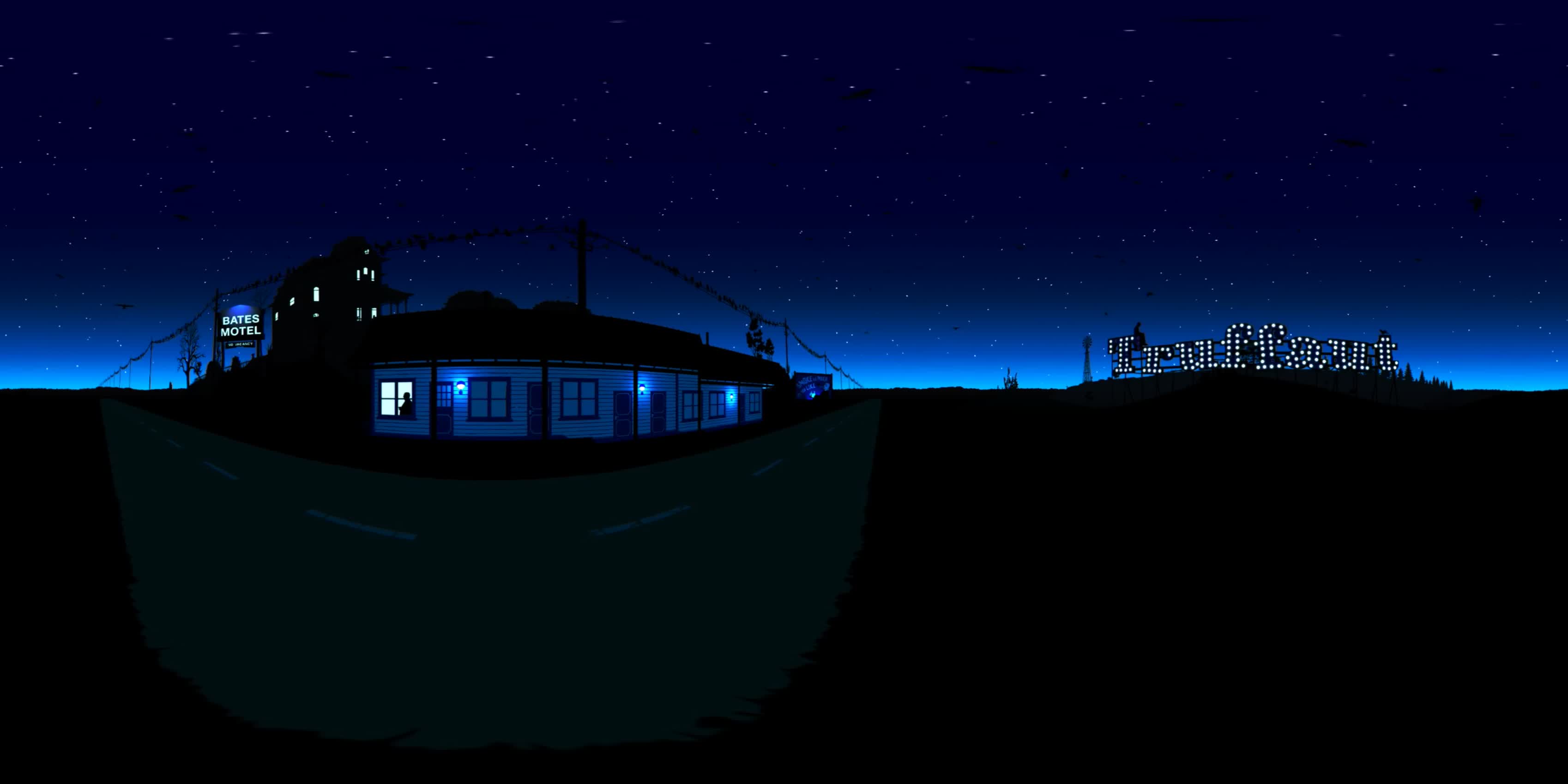} \vspace{-1em} \\

A Fisherman’s Tale & 5:20 & 
A fisherman puppet who is trapped inside a model lighthouse finds itself inside an infinite loop. & 
\vspace{-1em} \includegraphics[width=\linewidth]{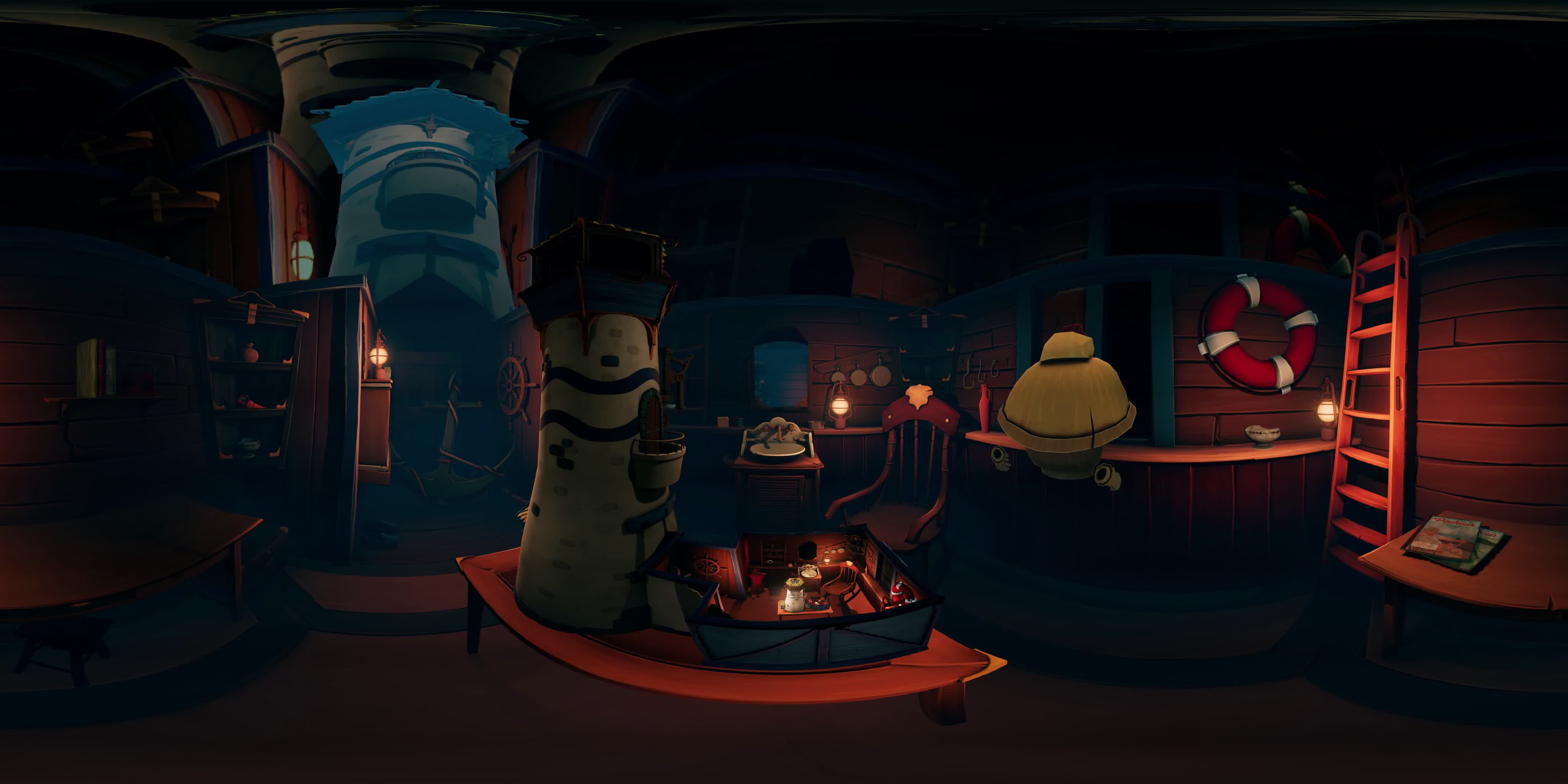} \vspace{-1.5em} \\

\bottomrule
\end{tabular}
\end{table*}

\subsection{360-Degree Cinematic Stories}

We stimulus sampled five pieces of 360-degree cinematic stories of comparable length to increase generalizability and minimize content-specific influences (Table \ref{tab:content}).\footnote{This deviated from our pre-registration of sampling three stimuli across experiments. The adjustments ensured that no participants repeated stimuli.} The content selection process was informed by pilot study data of individuals watching 10 pieces of publicly available 360-degree stories and guided by the following criteria. To control for story duration, we first filtered out content that was substantially longer or shorter than the others. We then removed content that is designed explicitly for a single user (e.g., content viewed from a passenger seat). Finally, we excluded content that were horror-themed and those that contained no virtual characters.

Our choice of monoscopic 360-degree stories was guided by three considerations. First, they remain one of the most prominent forms of VR content that leverages the medium's unique affordances. In contrast to 2D screen-based co-watching, where content is projected onto a virtual plane and viewed passively, 360-degree stories place viewers inside the scene and give them agency over what they see through head rotation. Second, we excluded stereoscopic 360-degree video, which is less widely produced and therefore difficult to sample stimuli from; mismatches between the rendered and the viewer's interpupillary distance also caused discomfort and motion sickness during pilot testing. Third, access to high-fidelity tracking data in a networked setting was essential to our exploratory aims. ENGAGE supported this but constrained our content options: during early testing, we deemed 360-degree video to be the most feasible format, as volumetric video and more interactive 3D stories were considerably less reliable or unavailable in sufficient quantity.

Specifically, Experiments 1 and 2 sampled across the first three stories in Table \ref{tab:content}, and Experiment 3 sampled the latter two. For Experiments 1 and 2, each group was assigned to watch two of the three stories, with each story equally sampled across conditions. In these two experiments, we also randomized the story assigned to the pairs of groups in the same timeslot to avoid confounding stimuli with the time of day for the VR exercises. 
%Recall that in Section \ref{sec: participants}, we split participants assigned to the same timeslot into two groups to allow for manipulation of group familiarity in Experiment 3 by shuffling group membership across the pairs of groups. 
For Experiment 3, participants were shuffled into newly formed groups to manipulate group familiarity, which resulted in some groups where all three of the selected stories had been watched by at least one participant. To ensure that no participants repeated stimuli in Experiment 3, we further sampled two additional stories.

\begin{figure}[t]
 \centering 
 \includegraphics[width=0.4\textwidth,keepaspectratio]{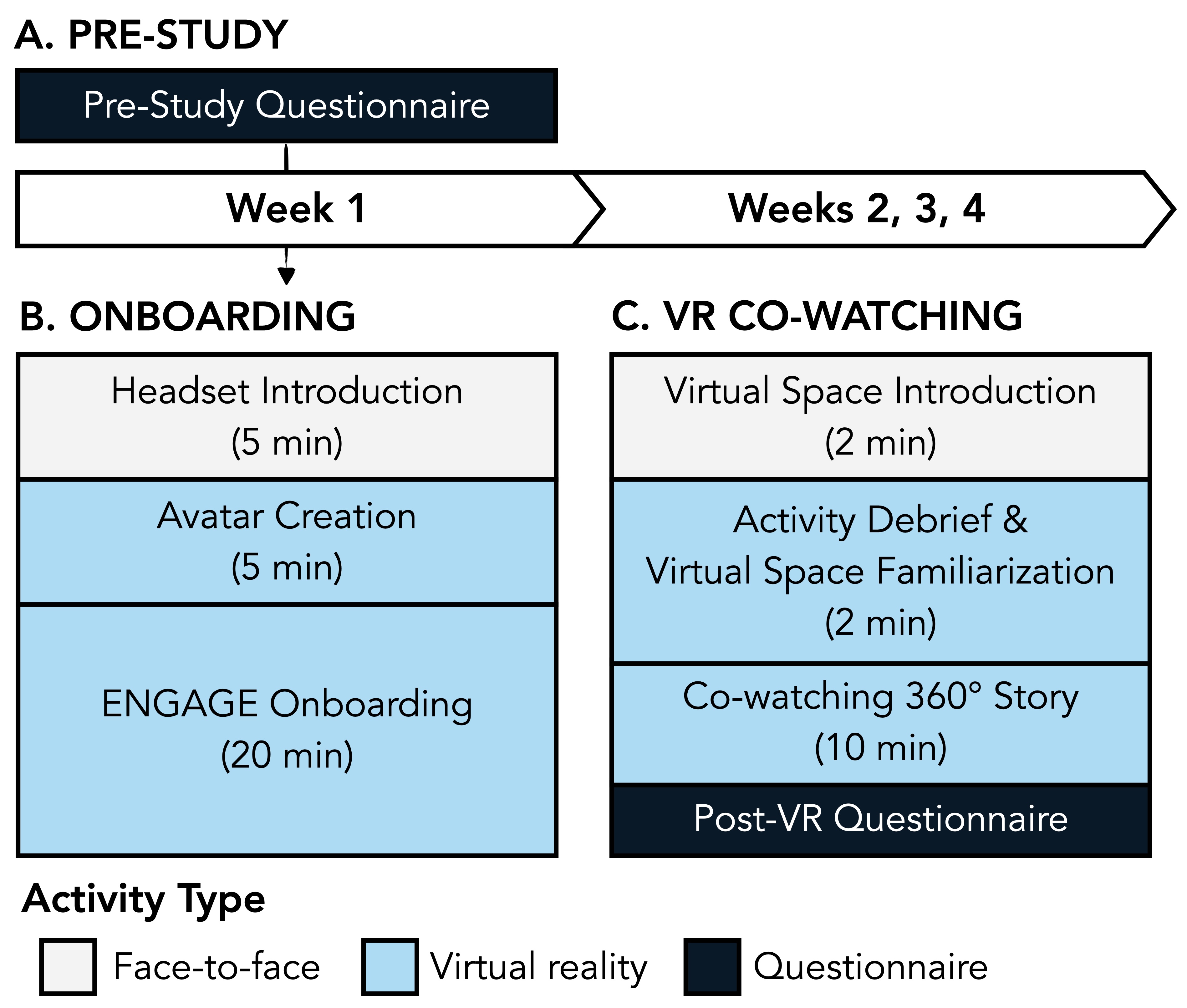}
 \caption{Overview of the Four-Week Study Procedure. 
 %The protocol for VR co-watching was repeated across weeks 2, 3, and 4. Across the final three weeks, we studied the effects of spatial boundary type, size of personal space, and group familiarity.
}
 \label{fig: procedure timeline}
\end{figure}

\subsection{Procedure}
\label{sec: procedure}
Figure \ref{fig: procedure timeline} outlines our four-week study procedure, which consists of a pre-study stage, onboarding, and the three experiments. In the pre-study stage, participants completed a consent form and a questionnaire on demographics (Figure \ref{fig: procedure timeline}A). %Availabilities were also collected in the questionnaire, which allowed the third-party coordinator to assign participants to groups as detailed in Section \ref{sec: participants}.

In week 1, participants gathered with their group in a classroom to complete onboarding (Figure \ref{fig: procedure timeline}B). Following headset introduction led by a researcher, participants donned Meta Quest 2 headsets and logged into ENGAGE. They were then guided to spend five minutes creating an avatar that looks and feels like them using the platform’s avatar creation tool, which allowed for customizations of an avatar’s physical attributes (e.g., hairstyle, skin tone), outfits, and accessories. Participants wore the same avatar in all remaining VR sessions. During the VR sessions, participants embodied half-body avatars, as using full-body avatars overlaid on 360-degree stories greatly increased visual incongruency. Avatar head position and rotation were driven by the headset's 6DOF tracking data (i.e., position, rotation), and hand poses and gestures by the 6DOF tracking of the two controllers. Each avatar is further identified through a name tag positioned above their avatar. Because the Quest 2 headset does not support eye tracking, gaze and attention can only be conveyed through avatar head direction. Participants were co-located in the same physical room, so their speech carried directly between them; we therefore muted their microphones in ENGAGE to prevent echo, which meant that avatars did not lip-sync while participants spoke. After this, the participants joined a private VR session, and were given instructions on how to move (i.e., smooth translation, teleportation, physical movement) and look around (i.e., head turning, yaw rotation using controller) in virtual environments. The participants also spent time interacting with each other, completing ice-breaking exercises such as taking group selfies.

In week 2, participants met in the same physical classroom as their group (Figure \ref{fig: procedure timeline}C). The session took around 30 minutes, with approximately 12 spent in VR; participants were seated in swivel chairs and allowed to move freely, while being mindful of others. They were first debriefed on the activity, shown an overview of the spatial setup of the virtual environment associated with their assigned condition, and entered into an ENGAGE session with their group. The virtual environment was instantiated with an IFXSnapshot containing ground markers to guide the participants. Once all participants joined the session, the researcher asked the participants to move to the ground markers. As outlined in Figure \ref{fig: conditions}A and later detailed in Section \ref{sec: procedure}, participants in the reserved seating condition were instructed to move individually to a 1×1 meter ground marker, while those in the box and general admission spatial conditions moved to a communal 20×20 meter marker. For configurations with individual markers, participants picked their own marker, mirroring real-world seating behavior and avoiding the overhead of assigning positions to as many as ten users. The researcher then added the invisible boundaries as an IFXSnapshot, and deleted the visible ground markers to complete the virtual spatial setup. This procedure ensured that participants were properly positioned within the invisible boundaries. Participants then spent one minute familiarizing themselves with the virtual space. Following this, a researcher reiterated the goal of the exercise and began the 360-degree story using the platform’s video player. When the story ended, participants exited VR and completed a post-VR questionnaire. 

\begin{figure*}[t]
 \centering 
 \includegraphics[width=.95\textwidth,keepaspectratio]{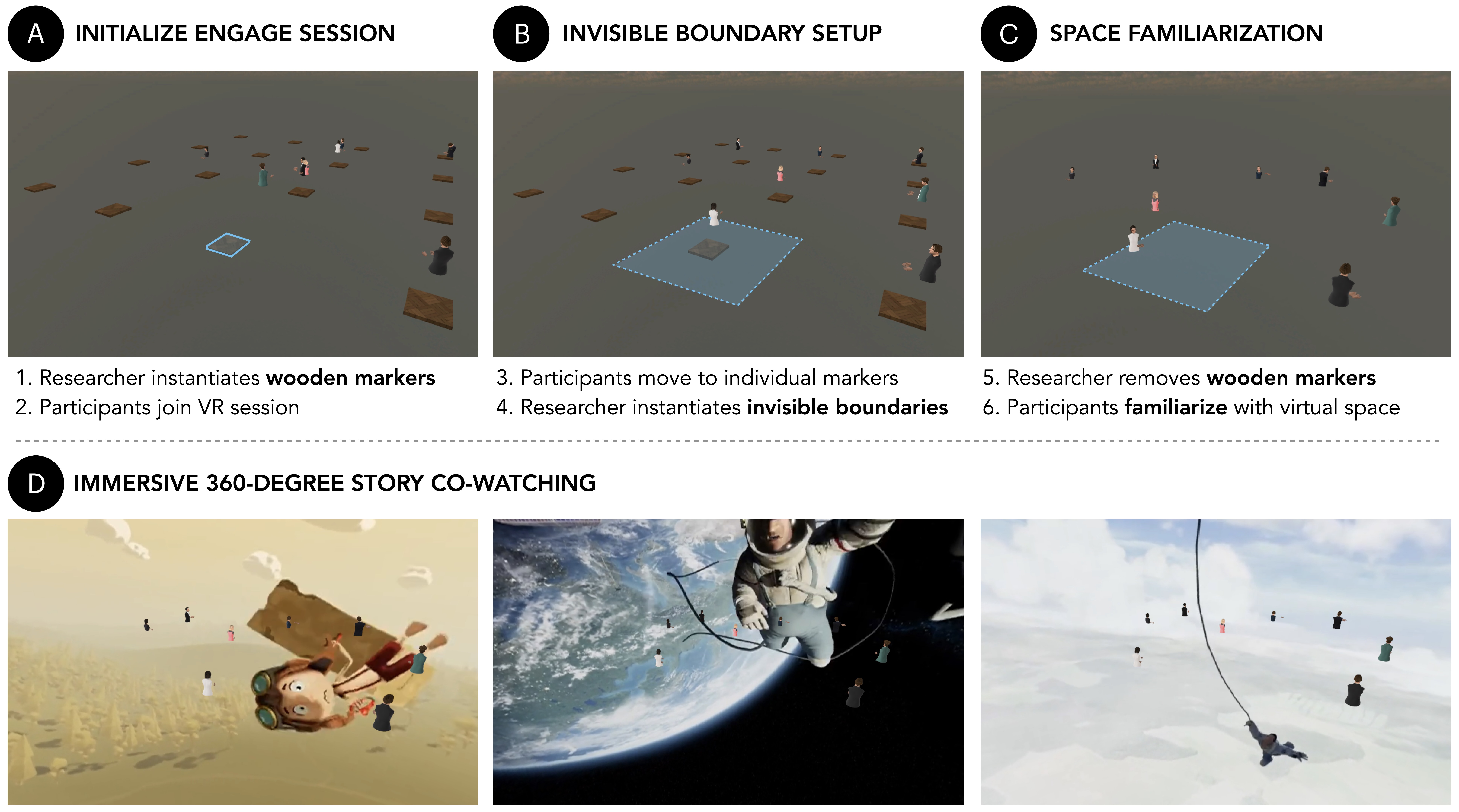}
 \caption{ENGAGE Screenshots of Group assigned to Reserved Seating condition. In subfigures a–c, the screenshots denote study protocols for setting up the invisible boundaries for individual users, where the blue overlays and outlines highlight the boundaries and wooden marker eventually occupied by the user in the white top. Subfigure d shows exemplar screenshots from three moments of exercise from the group watching ARIPI. Name tags appeared above each avatar during the experiments but have been hidden in these screenshots to preserve anonymity.}
 \label{fig: engage}
\end{figure*}

The protocol in week 2 was repeated in weeks 3 and 4, with the virtual spatial configuration set according to the assigned condition for each group (Figure \ref{fig: engage}). In week 4, for the groups that were assigned to the shuffled condition, we shuffled the group assignments across the pairs of concurrently running groups to manipulate group familiarity, so participants met and gathered with their updated groups in separate physical rooms to complete the VR co-watching exercises.

\subsection{Self-Report Measures} \label{sec: self-report measures}
This section describes measures collected pre- and post-study after each experiment. All post-study measures were collected after each experiment, except group familiarity, which was measured only at the end of Experiment 3.

\textbf{Demographics}. Participants reported their age, gender, ethnicity, and prior XR experiences in the pre-study questionnaire.

\textbf{Presence}. We measured social and spatial presence, that is, the sense of being there with social others (i.e., users, story characters) and in the virtual environment, respectively. Social presence is measured using a 3-item 5-point Likert scale from “Not at all” to “Extremely” following \cite{Han2023, Wang2024}. We measured social presence towards other users and character(s) in the immersive stories. The constructs had high reliability across experiments ($\alpha_{users}$=.84; $\alpha_{char.}$=.76), so were averaged to create two composite scores (M$_{users}$=2.18, SD$_{users}$=0.94; M$_{char.}$=2.10, SD$_{char.}$=0.88). We measured spatial presence using a 3-item 5-point Likert scale \cite{Han2023}, which was also averaged to create one score ($\alpha$=0.78, M=2.69, SD=0.89).

\textbf{Perceived Attention}. Participants reported their attention towards the story and social others. Attention towards the story was measured using the 3-item attentional focus construct in the Narrative Engagement Scale \cite{Busselle2009Measuring}. We adapted the original items, replacing ``program'' with ``360-degree story'', using a 7-point scale, from ``Strongly disagree'' to ``Strongly agree.'' The items were averaged to create one score ($\alpha$=.91, M=3.59, SD=1.53). Attention towards social others was measured using a 3-item 7-point Likert scale based on the attentional allocation construct in \cite{Harms2004}. The top three items were selected based on factor loading scores. The items had low reliability ($\alpha$=.22), so we examined them individually, focusing on the two items that measured a user’s attention towards others (M=2.65, SD=1.39) and the attention they perceived from others (M=1.78, SD=1.55). Because single-item measures tend to have lower construct validity than multi-item measures, we interpret results with caution, as null findings may reflect measurement noise rather than the absence of an effect.
    
\textbf{Workload}. We measured workload using the NASA-TLX scale \cite{Hart1988}. We selected four items from the original scale that were most relevant: mental demand (M=2.57, SD=1.40), perceptual demand (M=3.05, SD=1.57), physical demand (M=1.73, SD=1.07), and frustration (M=3.06, SD=1.44), with each item measured on a 7-point scale. 
    
\textbf{Motion Sickness}. Motion sickness was assessed using the 9-item, 4-point Virtual Reality Sickness Questionnaire \cite{Kim2018}, from ``Not at all'' to ``Very.'' We derived scores for oculomotor (M=23.42, SD=19.98) and disorientation (M=15.29, SD=16.88) following \cite{Kim2018}, each scaled from 0 to 100.
    
\textbf{Group Familiarity}. Following \cite{Castella2000, Gruenfeld1996}, familiarity was measured using a one-item 5-point Likert scale from ``Not at all well'' to ``Extremely well'' (M=1.73, SD=0.71).

\subsection{Behavioral Measures}\label{sec: behavioral measures}
We outline behavioral data captured from VR session recordings, trimmed to time periods when the story was playing. Data processing was done in Python.

\textbf{Interpersonal Distance with All Users}. We calculated interpersonal distance between each user and their group, averaging it over the story duration. The construct is measured in meters (M=11.56, SD=6.52) and signals spatial dynamics, as well as relationship closeness and social presence \cite{Hall1966, Oh2018}. It also reflects diegetic consistency of the experience, as extreme interpersonal distances (e.g., users standing too far) can perceptually conflict with the story’s spatial context. 

\textbf{Virtual Movement}. We derived the amount of virtual movement for each individual in the virtual environment. Like interpersonal distance, virtual movement provides insights on how individuals manage the social context and engage in spatial negotiation. This measure was calculated by aggregating the physical movement of the user’s head and virtual locomotion (i.e., smooth translation, teleportation) in 3D space. Total movement is divided by story duration, yielding a unit of meters per second (M=0.14, SD=0.22).
    
\textbf{Users in Field of View}. We derived the average number of user avatars in an individual’s field of view (FOV). Because the Meta Quest 2 headset does not track gaze, this measure approximates visual attention and captures the extent to which co-watchers were visually salient during the story. Prior work has similarly operationalized attention by examining the visual content in one's FOV \cite{Castellana2025, Miller2023}. Importantly, the presence of other users in FOV can introduce perceptual inconsistencies between the story world and the virtual environment occupied by the social group, whereas a co-watcher outside the FOV would not produce such effects. To calculate this, we extracted the virtual head positions and rotations (i.e., pitch, yaw). For each user, we then aggregated the number of users whose virtual head position falls within the given user’s FOV (i.e., 104 degrees horizontal, 98 degrees vertical), and averaged this value over the story duration (M=2.30, SD=1.19). 

\begin{table*}[t]
\centering
\small
\caption{Overview of Significant Self-Report Findings. Significant contrasts are shown (*: $p$$\leq$.05, **: $p$$\leq$.01, ***: $p$$\leq$.001). GA=General Admission; RS=Reserved Seating.}
\label{tab: Finding Summaries}
\renewcommand{\arraystretch}{1.05}
\setlength{\tabcolsep}{3pt}
\begin{tabular}{lll}
\toprule
\textbf{Outcome Variable} & \textbf{Predictor} & \textbf{Significant Contrasts} \\
\midrule

\multicolumn{1}{l}{\textbf{Presence}} \\
\hspace{0.15cm}Social (Users) & Spatial Boundary Type ($p$=.0050) & Box > GA*, Box > RS*\\
&Size of Allocated Personal Space ($b$=-0.070, $SE$=0.034, $p$=.041) & \\
\hspace{0.15cm}Social (Character) & Group Familiarity ($b$=0.24, $SE$=0.11, $p$=.027) & \\
\hspace{0.15cm}Spatial & Group Familiarity ($b$=0.34, $SE$=0.12, $p$=.0048) & \\

\addlinespace

\multicolumn{1}{l}{\textbf{Perceived Attention}} \\
\hspace{0.15cm}From Others &\emph{No predictor was significant} & \\
\hspace{0.15cm}Towards Others &\emph{No predictor was significant} & \\
\hspace{0.15cm}Towards Story & Group Familiarity ($b$=0.45, $SE$=0.22, $p$=.043) & \\

\addlinespace

\multicolumn{1}{l}{\textbf{Workload}} \\
\hspace{0.15cm}Mental Demand & Spatial Boundary Type ($p$=.0081) & GA > RS*, Box > RS*\\
\hspace{0.15cm}Perceptual Demand & Group Familiarity ($b$=0.47, $SE$=0.23, $p$=.044) & \\
\hspace{0.15cm}Physical Demand & Group Familiarity ($b$=0.35, $SE$=0.14, $p$=.019) & \\
\hspace{0.15cm}Frustration & Size of Allocated Personal Space ($b$=-0.11, $SE$=0.047, $p$=.019) & \\

\addlinespace

\multicolumn{1}{l}{\textbf{Motion Sickness}} \\
\hspace{0.15cm}Oculomotor & \emph{No predictor was significant} & \\
\hspace{0.15cm}Disorientation & Size of Allocated Personal Space ($b$=-1.07, $SE$=0.49, $p$=.030) & \\

\addlinespace

\multicolumn{1}{l}{\textbf{Behavioral}} \\
\hspace{0.15cm}Virtual Movement & Spatial Boundary Type ($p$<.001) & GA > RS**, Box > RS***\\

\hspace{0.15cm}Interpersonal Distance & Spatial Boundary Type ($p$<.001) & GA > RS***, GA > Box***, RS < Box**\\
& Size of Allocated Personal Space ($b$=0.14, $SE$=0.0072, $p$<.001) &\\
\hspace{0.15cm}Users in FOV & \emph{No predictor was significant} & \\
\hspace{0.15cm}FOV Covered by Users & Spatial Boundary Type ($p$<.001) & Box > GA***, Box > RS***\\
& Size of Allocated Personal Space ($b$=-0.38, $SE$=0.040, $p$<.001) &\\

\bottomrule
\end{tabular}
\end{table*}

\textbf{Field of View Covered by Users}. We calculated the percentage that one’s FOV was occupied by other user avatars. This measure extends the previous one by quantifying not only whether co-watchers were visible, but how much of the visual field they occupied. In other words, while the previous measure is agnostic to interpersonal distance, the percentage of FOV covered by users is not: a co-watcher standing closer in one’s FOV will yield higher coverage compared to one standing further away. As such, the measure captures both social attention and the degree to which co-watchers occluded the immersive story content, providing an indicator of diegetic consistency and perceptual interference.

For a given user, the positions of other users were projected into the given user’s angular view based on head position and rotation (yaw, pitch, and roll), assuming again the FOV of the Meta Quest 2 headset (104-degree horizontal, 98-degree vertical). Consistent with ENGAGE’s approach in rendering virtual avatars, we only included avatars located in front of the user and beyond a clipping distance of 0.2 meters. The amount of FOV covered by a visible avatar is approximated as upright rectangular silhouettes whose angular dimensions are scaled with distance. The silhouette dimension for avatars was set to 1 meter tall and 0.3 meter wide, which were estimated using virtual camera captures and referenced against a 1-meter cube. To avoid double-counting overlapping visual regions across avatars, we computed the union of avatar regions within the FOV and divided the resulting non-overlapping angular area by the total FOV area to obtain a percentage. This measure was finally averaged across the story duration (M=1.48, SD=1.44).

\subsection{Data Analysis}\label{sec: data analysis}
To predict self-report outcomes, we used linear models with condition as the predictor. We tested linear mixed-effects models with group as the random effect, but likelihood ratio tests found that almost all models with the added parameterization did not significantly improve model fit. We also tested models variations where content (i.e., story shown) and group size are used as predictors, but they were non-significant in predicting the majority of the outcomes. As such, we chose the more parsimonious linear model with condition as the predictor. 

For predicting behavioral measures, we used linear models as the more complicated linear mixed-effects models with group added as random effect did not yield significantly better fit using likelihood ratio tests. We again tested group size and content as possible predictors and found that while content did not significantly predict most outcomes across experiments, group size did. The final model was therefore a linear model with condition and group size as its predictors. For the measures of interpersonal distance and the amount of virtual movement, we log-transformed them as their raw values violated the normality assumptions of linear models.

Furthermore, spatial configuration in Experiment 1 was modelled as a categorical variable, while Experiments 2 and 3's operationalization of reserved seating area size and group familiarity was treated as continuous measures of the dimension of the bounded area and self-report group familiarity. In Experiment 1, where the manipulation of spatial configuration is at three levels, we report post-hoc pairwise comparisons with Bonferroni correction. We built all models using R. Data points falling outside three standard deviations from the mean are winsorized, and significance is determined at $\alpha$=.05. Table \ref{tab: Finding Summaries} summarizes key findings across the experiments.

\section{Experiment 1: Type of Spatial Boundary}\label{sec: exp 1}

\subsection{Demographics}
Participants (Female=67, Male=42) between ages 18 and 29 (M=20.78, SD=1.58) identified as Asian or Asian-American (N=52), Caucasian (N=24), Hispanic or LatinX (N=10), African, African-American or Black (N=9), Multi-racial (N=6), Middle Eastern (N=3), Native Hawaiian or other Pacific Islander (N=2), or declined to answer (N=2). Participants varied in their previous XR experiences (None=22, One to two times=45, Three to five times=24, At least six times=18).

\begin{figure*}[tbp]
 \centering 
 \includegraphics[width=0.7\textwidth,keepaspectratio]{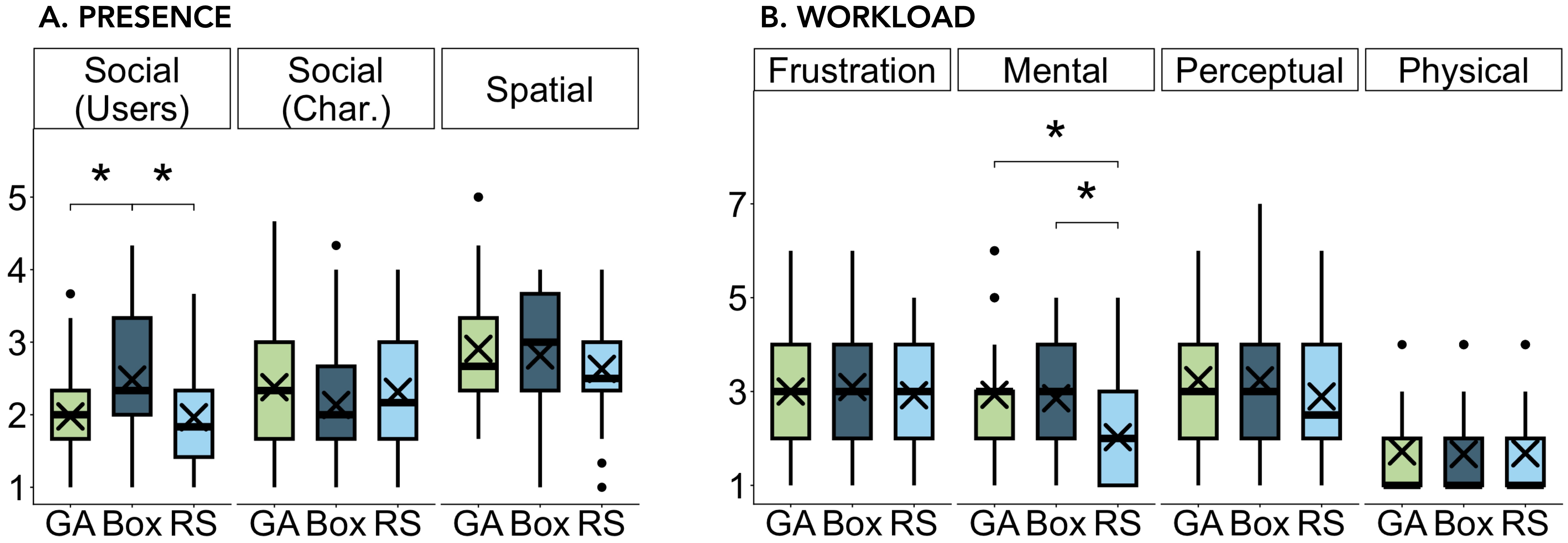}
 \caption{Experiment 1 Self-Report Findings. Significant contrasts are shown (*: $p$$\leq$.05, **: $p$$\leq$.01, ***: $p$$\leq$.001). Char.=Character(s). GA=General Admission; RS=Reserved Seating.}
 \label{fig: exp1 selfreport}
\end{figure*}

\begin{figure}[tbp]
 \centering 
 \includegraphics[width=0.3\textwidth,keepaspectratio]{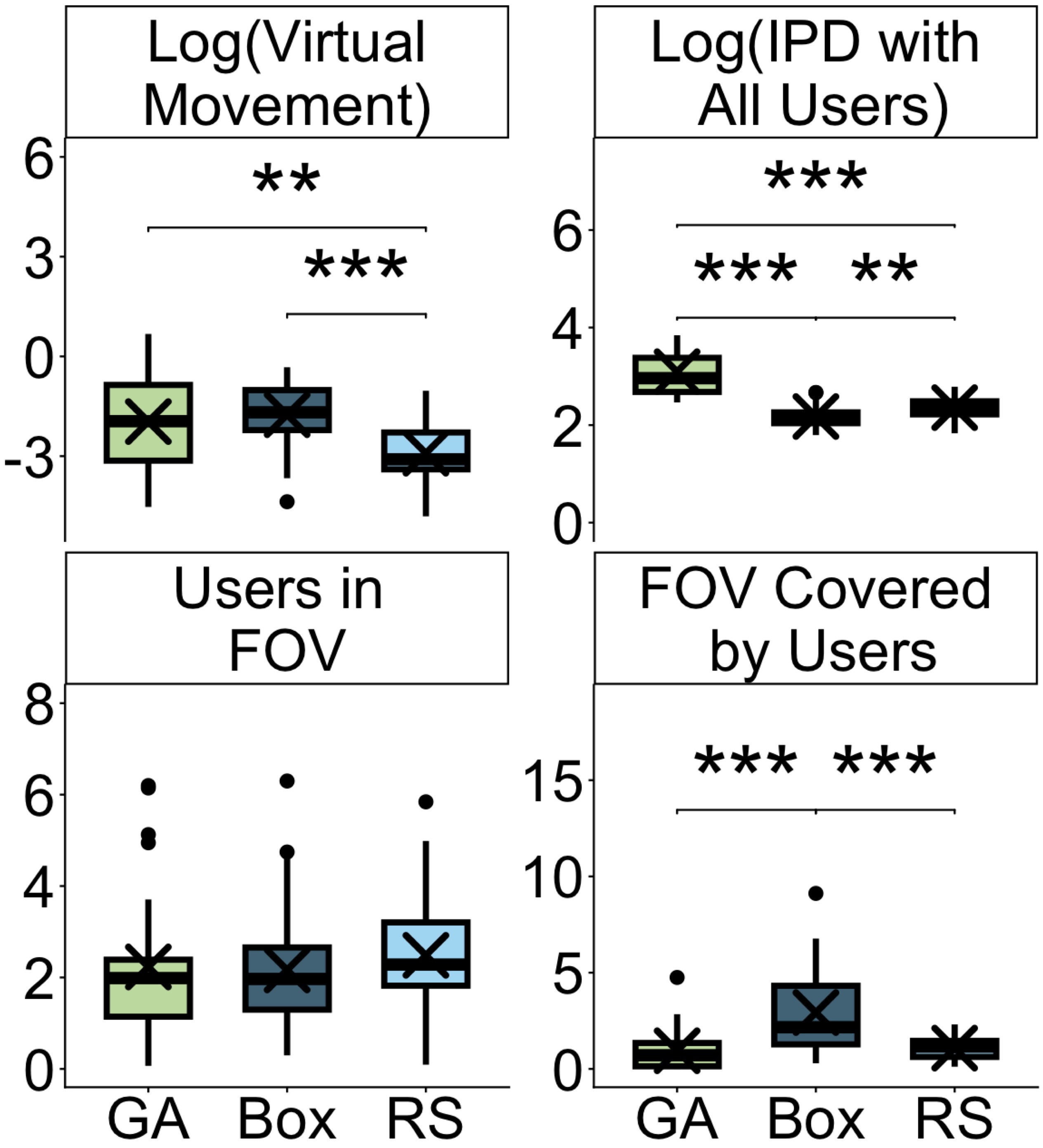}
 \caption{Experiment 1 Behavioral Findings. Significant contrasts are shown (*: $p$$\leq$.05, **: $p$$\leq$.01, ***: $p$$\leq$.001). GA=General Admission; RS=Reserved Seating; FOV=Field of View; IPD=Interpersonal distance.}
 \label{fig: exp1 behavioral}
\end{figure}

\subsection{Conditions}
To study \textbf{RQ1a}, the first experiment compared three types of virtual spatial boundaries during social co-watching (Figure \ref{fig: conditions}A). Informed by the naming conventions of seating arrangements used for social gatherings, such as in concerts and sports games, the three configurations are \textit{general admission}, \textit{box}, and \textit{reserved seating}. Groups of participants were randomly assigned one of the three conditions.

In the general admission condition, participants were allowed to explore the virtual environment with little spatial constraints. In practice, the default virtual environment used for showing 360-degree videos in ENGAGE has a spatial boundary of approximately 550 by 470 meters, which we further constrained into a square spatial boundary of 400 by 400 meters using invisible boundaries. In the box condition, participants are constrained by an invisible communal boundary of 20 by 20 meters. 

Finally, in the reserved seating condition, participants are individually constrained in a smaller square meter box of 4 by 4 meters using invisible boundaries. The individual reserved seating areas are lined up in a grid-like fashion with equal spacing (i.e., 2m) between one another (see Figure \ref{fig: conditions} for schematics of the grid). The choice of two meters aligns with the appropriate interpersonal distance for social spaces per the Proxemics Theory \cite{Hall1966}. We chose a grid-based configuration over alternatives common seen in shared spaces, such as rows and circles, for two reasons. First, unlike rows, a grid assumes no forward direction, which suits 360-degree stories that often lack one. Second, grid-based configurations scale more cleanly across group sizes than circular arrangements do, which mattered because our groups ranged from five to ten participants and attendance varied week to week. We piloted circular configurations and found that partially filled circles produced disparate levels of avatar occlusion and interpersonal distance within the same configuration, and made interpersonal distance and size of allocated space more difficult to operationalize compared to a grid-based design.

Notably, the reserved seating condition differs from the box and general admission conditions in two dimensions: boundary type (i.e., individual, communal) and size of spatial constraint (e.g., 4x4m, 400x400m). This confound makes it difficult to disentangle each dimension's individual effect. Our reason for doing so is twofold. To start, the three conditions are grounded in real-world use cases as noted in Section \ref{section: motivation for spatial configurations}, as crossing the two dimensions would yield configurations with no real-world analogue (e.g., individual 400x400m boundaries). Additionally, this design allows us to explore multiple dimensions before delving into specific manipulations in later experiments. Experiment 1 is better read as a comparison of three ecologically grounded configurations than as a test of any one spatial manipulation. That said, the box and general admission conditions hold boundary type constant, allowing us to examine the effect of spatial constraint size.

\begin{figure}[tbp]
 \centering 
 \includegraphics[width=0.3\textwidth,keepaspectratio]{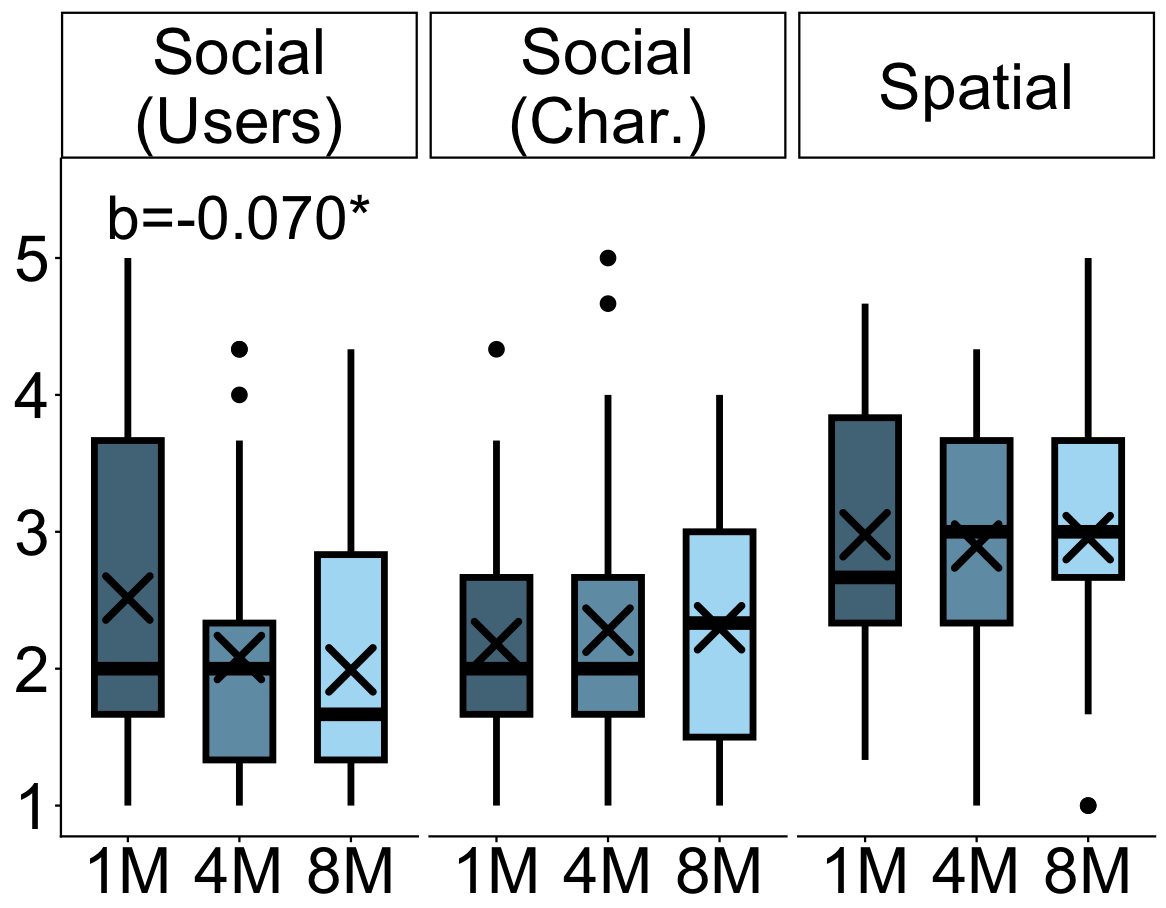}
 \caption{Experiment 2 Presence Findings. Significant effects of personal space size are shown (*: $p$$\leq$.05, **: $p$$\leq$.01, ***: $p$$\leq$.001). Char.=Character(s).}
 \label{fig: exp2 presence}
\end{figure}

\subsection{Results}
\subsubsection{Self-Report Differences Across Spatial Boundary Types}
Figure \ref{fig: exp1 selfreport} summarizes self-report findings. Social presence towards others differed significantly across conditions ($p$=.0050), with those in box reporting higher scores compared to general admission ($b$=0.51, $SE$=0.19, $p$=.025) and reserved seating ($b$=0.52, $SE$=0.18, $p$=.011). Boundary type did not significantly predict social presence towards story characters ($p$=.45) and spatial presence ($p$=.37). It also did not significantly predict attention towards the story ($p$=.63), and the one-item measures on attention towards other users ($p$=.31), perceived attention from other users ($p$=.26), as well as the motion sickness measures of disorientation ($p$=.47) and oculomotor ($p$=.97). For workload, conditions significantly predicted mental demand ($p$=.0081), with those in reserved seating reporting lower scores compared to box ($b$=0.83, $SE$=0.30, $p$=.021) and general admission ($b$=0.90, $SE$=0.33, $p$=.023). The models revealed no significant differences for the remaining workload measures, namely frustration ($p$=.86), perceptual demand ($p$=.56), and physical demand ($p$=.97). 

\subsubsection{Behavioral Differences Across Spatial Boundary Types}

Figure \ref{fig: exp1 behavioral} shows our behavioral findings, which we do not regard as manipulation checks, as spatial configurations only afforded different mobility patterns but did not mandate them. 
%Furthermore, participants were also given the same instructions prior to the co-watching experience and were not explicitly guided to spatially position themselves in one way or another. 
The amount of virtual movement differed significantly across conditions ($p$<.001) and not group size ($p$=.66). There were significant contrasts between reserved seating and the conditions of box ($b$=1.13, $SE$=0.28, $p$<.001) and general admission ($b$=0.95, $SE$=0.29, $p$=.0042), with reserved seating exhibiting less movement. Interpersonal distance was significantly predicted by condition ($p$<.001) and group size ($b$=-0.066, $SE$=0.025, $p$=.0090), with significant contrasts between box and reserved seating ($b$=0.23, $SE$=0.072, $p$=.0051), box and general admission ($b$=0.97, $SE$=0.075, $p$<.001), and reserved seating and general admission ($b$=0.74, $SE$=.076, $p$<.001). Those in general admission stood furthest away from their group, followed by reserved seating, then box. The number of users in FOV was not significantly predicted by conditions ($p$=.58), but was for group size ($b$=0.28, $SE$=0.11, $p$=.013). User FOV coverage differed significantly across conditions ($p$<.001) and group size ($b$=0.26, $SE$=0.12, $p$=.031), with significant contrasts between box and general admission ($b$=2.20, $SE$=0.37, $p$<.001), and box and reserved seating ($b$=2.05, $SE$=0.35, $p$<.001), with box exhibiting higher values.

\section{Experiment 2: Size of Allocated Personal Space}
\label{sec: exp 2}

\begin{figure*}[tbp]
 \centering 
 \includegraphics[width=0.7\textwidth,keepaspectratio]{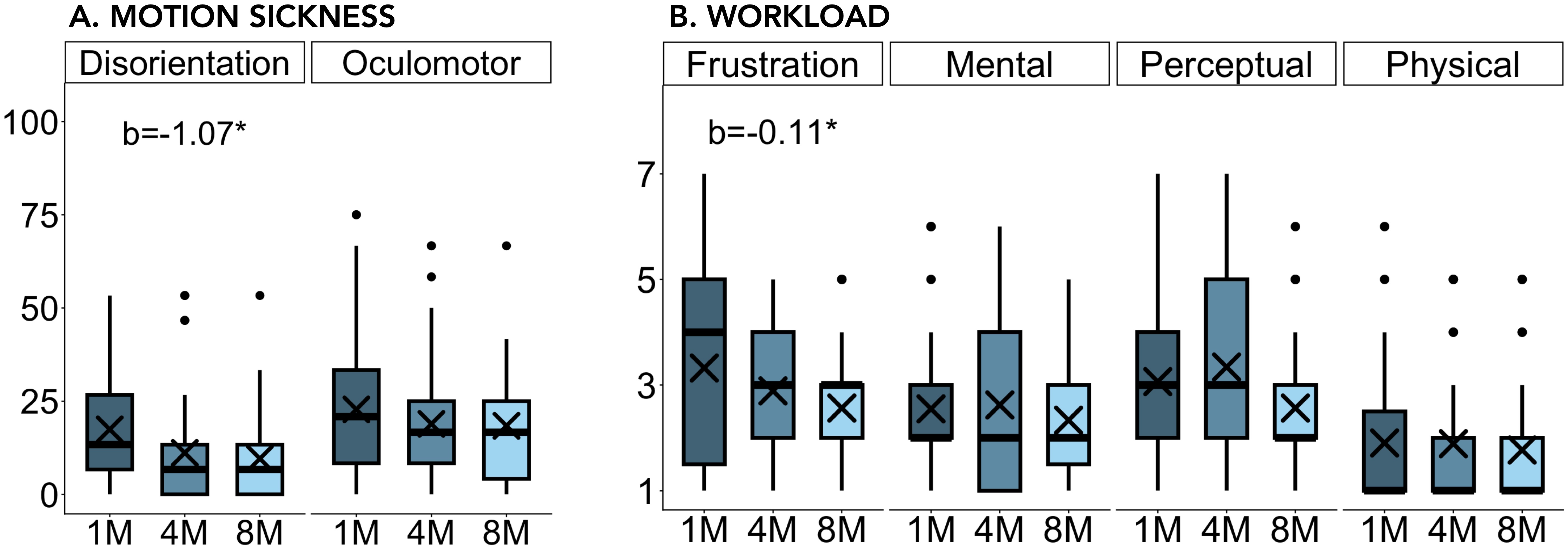}
 \caption{Experiment 2 Motion Sickness and Workload Findings. Significant effects of personal space size are shown (*: $p$$\leq$.05, **: $p$$\leq$.01, ***: $p$$\leq$.001).}
 \label{fig: exp2 selfreport}
\end{figure*}

\begin{figure}[tbp]
 \centering 
 \includegraphics[width=0.3\textwidth,keepaspectratio]{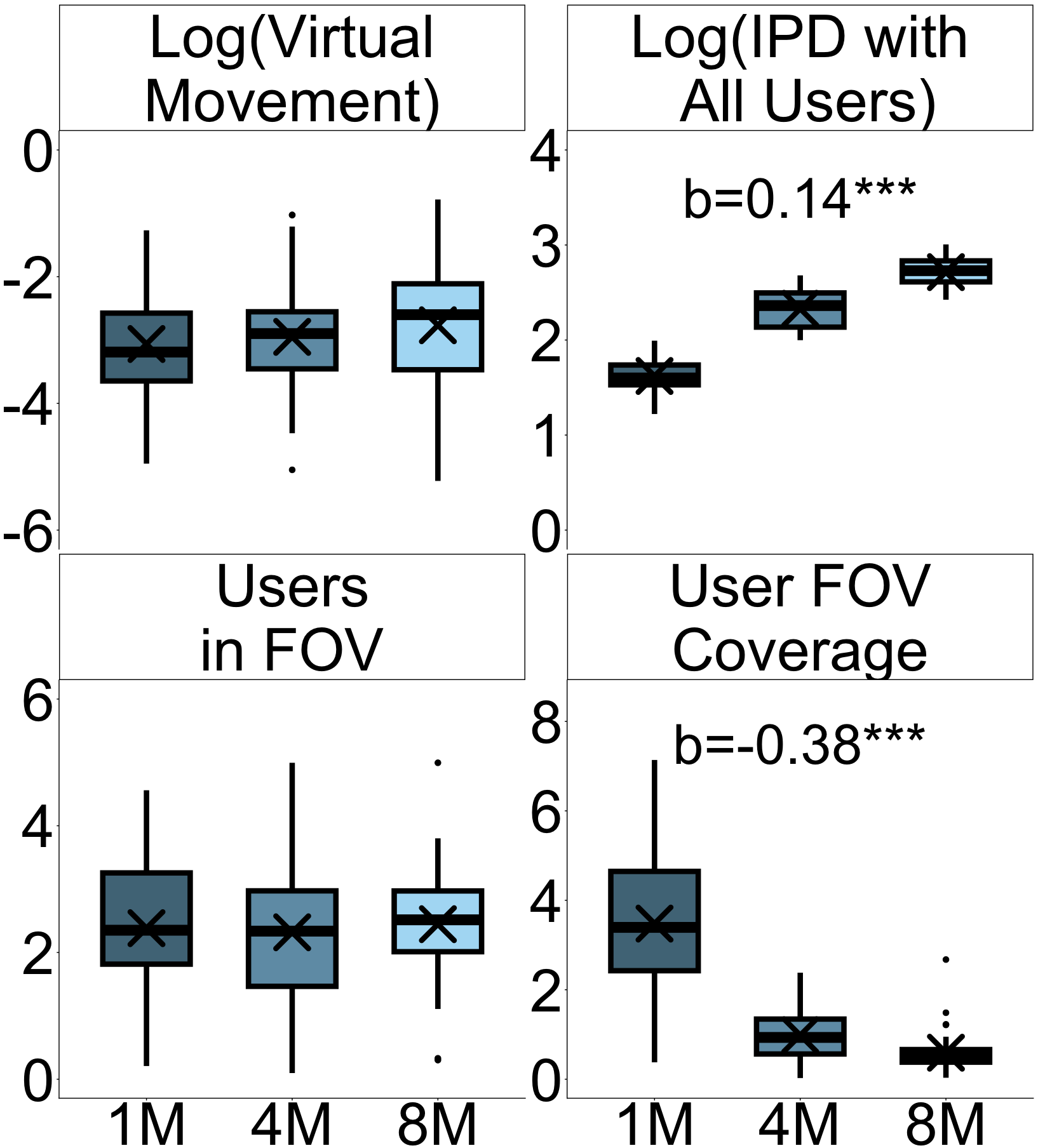}
 \caption{Experiment 2 Behavioral Findings. Significant effects of personal space size are shown (*: $p$$\leq$.05, **: $p$$\leq$.01, ***: $p$$\leq$.001). FOV=Field of View; IPD=Interpersonal distance.}
 \label{fig: exp2 behavioral}
\end{figure}

\subsection{Demographics}
Participants (Female=69, Male=47) between ages 18 to 29 (M=20.74, SD=1.58) identified as Asian or Asian-American (N=55), Caucasian (N=25), Hispanic or LatinX (N=10), African, African-American or Black (N=12), Multi-racial (N=7), Middle Eastern (N=2), Native Hawaiian or other Pacific Islander (N=3), Indigenous/Native American, Alaska Native, First Nations (N=1), or declined to answer (N=1). Participants varied in prior XR experiences (None=21, One to two times=49, Three to five times=25, At least six times=21). Because demographics were collected before Experiment 1, self-reported prior experiences do not reflect study participation.

\subsection{Conditions}
Experiment 2 investigates different instantiations of the reserved seating condition in Experiment 1 (\textbf{RQ1b}), specifically manipulating the size of personal space in each user’s individual reserved seating area (Figure \ref{fig: conditions}B). The experiment follows a 3x1 between-subjects design, where the size of the square meter box is set to 1x1, 4x4, and 8x8 meters. Distance between adjacent grids is set to 2 meters across all conditions, matching the reserved seating condition in Experiment 1. Groups were randomly assigned to one of the three conditions.

\subsection{Results}
\subsubsection{Self-Report Differences Across Personal Space Sizes}
Figures \ref{fig: exp2 presence} and \ref{fig: exp2 selfreport} summarize self-report findings. Participants in larger spaces reported lower social presence towards others ($b$=-0.070, $SE$=0.034, $p$=.041). Personal space size was non-significant in predicting social presence towards story characters ($b$=0.018, $SE$=0.030, $p$=.56) and spatial presence ($b$=0.0019, $SE$=0.030, $p$=.95). This was also the case for attention towards the story ($b$=0.042, $SE$=0.047, $p$=.37), and the one-item measures on attention towards other users ($b$=0.015, $SE$=0.051, $p$=.77), and from other users ($b$=-0.046, $SE$=0.055, $p$=.40). Those in larger personal space reported less disorientation ($b$=-1.07, $SE$=0.49, $p$=.030), while personal space size did not significantly predict oculomotor motion sickness ($b$=-0.64, $SE$=0.63, $p$=.31). Participants in smaller spaces reported greater levels of frustration ($b$=-0.11, $SE$=0.047, $p$=.019), though personal space size was not significant in predicting mental demand ($b$=-0.041, $SE$=0.047, $p$=.39), perceptual demand ($b$=-0.092, $SE$=0.053, $p$=.084), and physical demand ($b$=-0.023, $SE$=0.041, $p$=.57).

\subsubsection{Behavioral Differences Across Personal Space Sizes}
As shown in Figure \ref{fig: exp2 behavioral}, the amount of virtual movement did not differ significantly across sizes of personal space ($b$=0.043, $SE$=0.032, $p$=.19) and group size ($b$=0.0023, $SE$=0.076, $p$=.98). Interpersonal distance significantly increased with more personal space ($b$=0.14, $SE$=0.0072, $p$<.001) and in larger groups ($b$=0.054, $SE$=0.017, $p$=.0022). The number of users in FOV did not differ across personal space sizes ($b$=-0.010, $SE$=0.036, $p$=.77), yet was significantly predicted by group size ($b$=.25, $SE$=0.086, $p$=.0044). Coverage of other users in FOV was negatively predicted by the size of personal space ($b$=-0.38, $SE$=0.040, $p$<.001) and was not significantly predicted by group size ($b$=0.036, $SE$=0.094, $p$=.71).

\begin{figure*}[tbp]
 \centering 
 \includegraphics[width=0.85\textwidth,keepaspectratio]{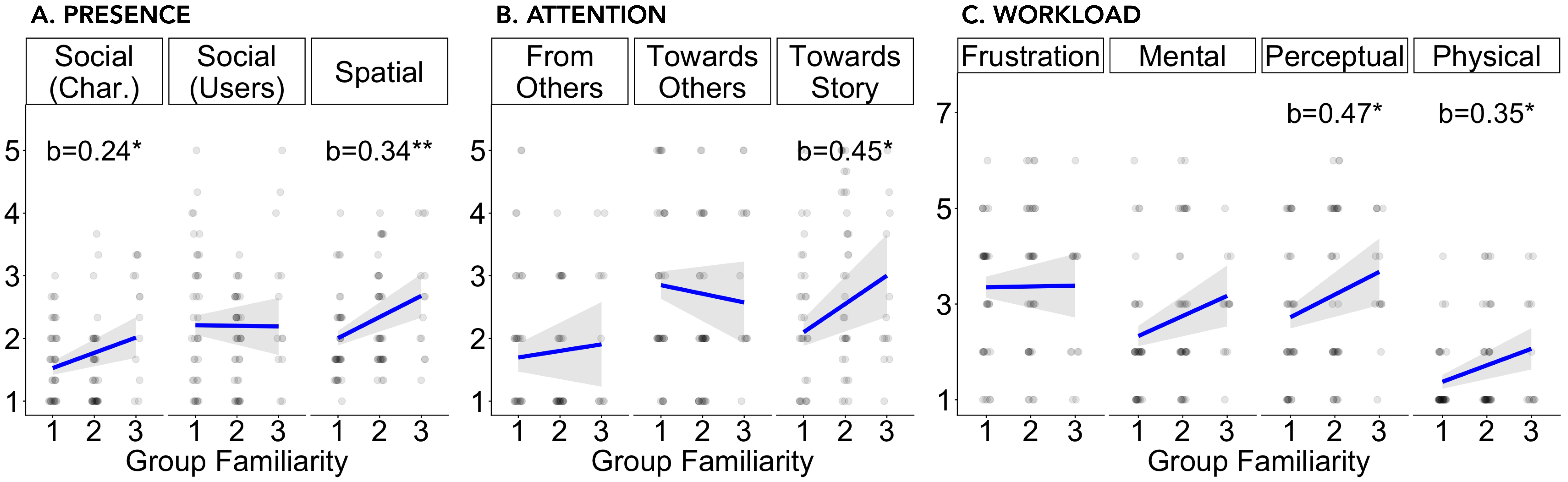}
 \caption{Experiment 3 Self-Report Findings. Blue lines represent model predictions for the prototypical user with mean group size of 8.06. Gray boundaries indicate predicted trajectories within one standard deviation of the modelled slope. Significant effects of group familiarity are shown. (*: $p$$\leq$.05, **: $p$$\leq$.01, ***: $p$$\leq$.001). Char.=Character(s).}
 \label{fig: exp3 selfreport}
\end{figure*}

\section{Experiment 3: Group Familiarity}\label{sec: exp 3}
\subsection{Demographics}
Participants (Female=56, Male=32) between ages 18 and 26 (M=20.73, SD=1.36) identified as Asian or Asian-American (N=42), Caucasian (N=18), Hispanic or LatinX (N=9), African, African-American or Black (N=8), Multi-racial (N=5), Middle Eastern (N=2), Native Hawaiian or other Pacific Islander (N=2), or declined to answer (N=2). Participants varied in prior XR experiences (None=16, One to two times=36, Three to five times=21, At least six times=15), which again did not reflect their study participation. 

\subsection{Conditions}
To study \textbf{RQ2}, we investigate group familiarity with a 2x1 between-subject design (Figure \ref{fig: conditions}C). In the shuffled condition, participants in the two groups assigned to the same time slot were shuffled into two newly assigned groups. In the non-shuffled condition, the groups are kept the same as in onboarding and the previous experiments. The two groups in time slots with only one group are placed in the non-shuffled condition, as shuffling is not feasible. We randomly assigned each remaining paired group one of the two conditions.

We aimed to induce differences in group familiarity by comparing non-shuffled groups, expected to exhibit higher familiarity, with shuffled groups, expected to exhibit lower familiarity. However, a manipulation check using a linear model to predict group familiarity with condition failed ($p$=.72), which we attribute to the lack of group bonding time despite this experiment occurring in the study's final week Subsequent analyses therefore used a self-reported 1-item 5-point group familiarity measure. Notably, one out of the 88 participants reported a group familiarity of ``Very well,'' and no participant reported knowing their group members ``Extremely well.'' This data point was treated as an outlier, as its value was beyond three standard deviations from the mean group familiarity and was winsorized to the next highest ordinal value.

%Specifically, we binned participants into those who reported the lowest group familiarity of “Not at all well” (N=38) and those who reported comparably higher levels of group familiarity (N=50), which we henceforth refer to as the \textit{low} familiarity and \textit{high} familiarity groups, respectively. 

\begin{figure}[tbp]
 \centering 
 \includegraphics[width=0.3\textwidth,keepaspectratio]{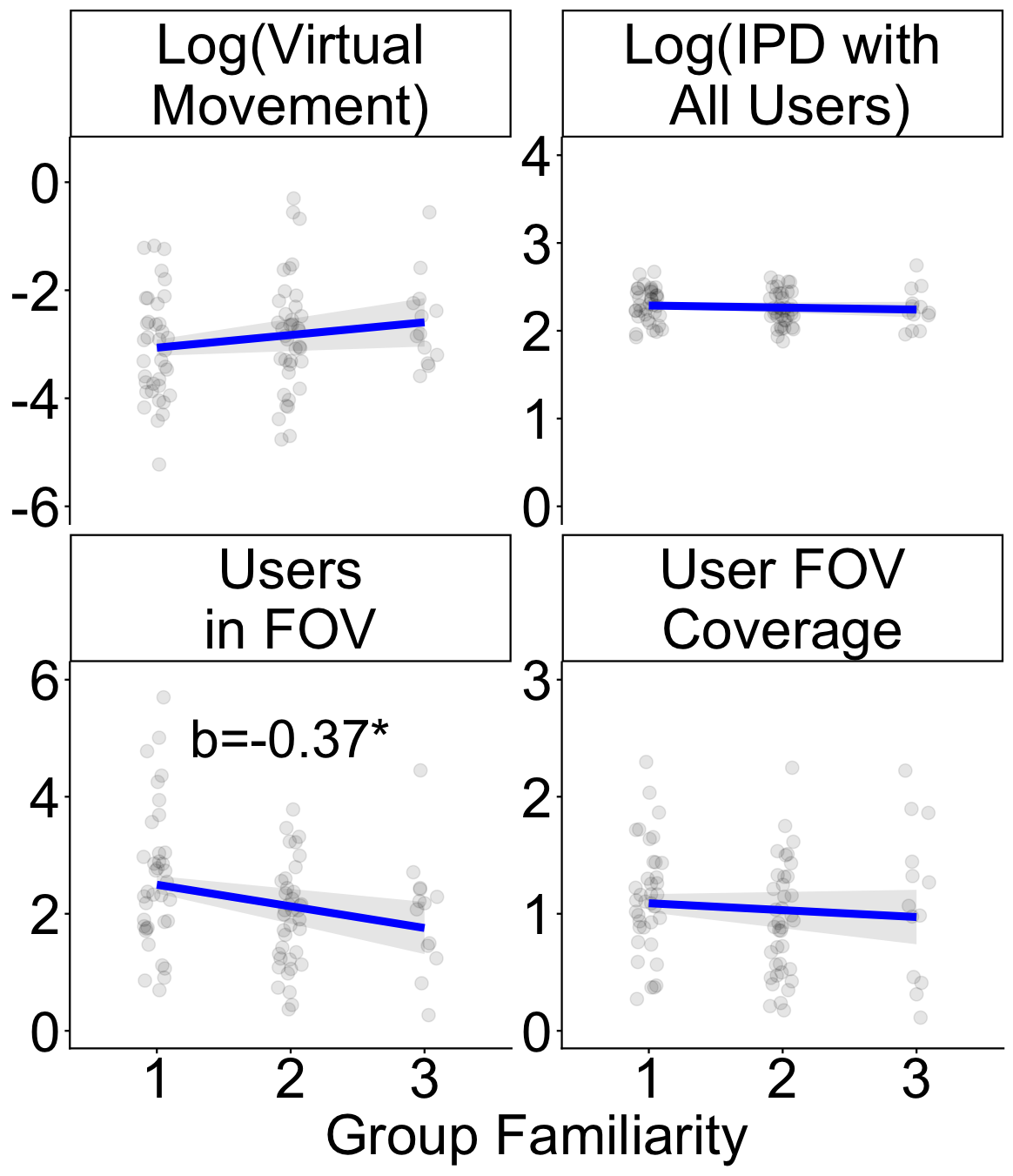}
 \caption{Experiment 3 Behavioral Findings. Blue lines represent model predictions for the prototypical user with the mean group size of 8.06. Gray boundaries indicate predicted trajectories within one standard deviation of the modelled slope. Significant effects of group familiarity are shown. (*: $p$$\leq$.05, **: $p$$\leq$.01, ***: $p$$\leq$.001). FOV=Field of View; IPD=Interpersonal distance.}
 \label{fig: exp3 behavioral}
\end{figure}

\subsection{Results}
\subsubsection{Self-Report Differences Across Group Familiarity}
We present findings on self-report measures in Figure \ref{fig: exp3 selfreport}. Group familiarity non-significantly predicted social presence towards users ($b$=-0.0098, $SE$=0.15, $p$=.95). It significantly predicted social presence towards story characters ($b$=0.24, $SE$=0.11, $p$=.027) and spatial presence ($b$=0.34, $SE$=0.12, $p$=.0048), with those with greater group familiarity indicating higher scores. There were no differences in the one-item measures of attention from others ($b$=0.10, $SE$=0.22, $p$=.64) and attention towards others ($b$=-0.14, $SE$=0.22, $p$=.53). Participants with higher group familiarity reported greater attention towards the story ($b$=0.45, $SE$=0.22, $p$=.043). Group familiarity did not significantly predict disorientation ($b$=2.48, $SE$=2.97, $p$=.41) and oculomotor motion sickness ($b$=2.16, $SE$=3.45, $p$=.53). For workload, participants with higher group familiarity scores indicated higher perceptual demand ($b$=0.47, $SE$=0.23, $p$=.044), and physical demand ($b$=0.35, $SE$=0.14, $p$=.019). Group familiarity did not significantly predict frustration ($b$=0.017, $SE$=0.22, $p$=.94) or mental demand ($b$=0.42, $SE$=0.21, $p$=.051).

%The model predicting social presence found no differences across group familiarity measures for both social presence towards users ($b$=-0.19, $SE$=0.21, $p$=.38) and towards the story characters ($b$=0.10, $SE$=0.16, $p$=.52). Spatial presence differed across low and high group familiarity, with those with high group familiarity reporting higher spatial presence than those reporting lower group familiarity ($b$=0.46, $SE$=0.17, $p$=.0064). Participants also differed in their attention towards the story, with those with higher group familiarity reporting higher levels of attention compared to those with lower group familiarity ($b$=0.68, $SE$=0.31, $p$=.031). There were no differences in the two other measures of attention, namely attention towards others ($b$=-0.41, $SE$=0.31, $p$=.18) and perceived attention from others ($b$=0.14, $SE$=0.32, $p$=.67).

\subsubsection{Behavioral Differences Across Group Familiarity}
Figure \ref{fig: exp3 behavioral} summarizes behavioral differences across group familiarity. For the amount of virtual movement, there were non-significant effects for group familiarity ($b$=0.23, $SE$=0.15, $p$=.12) and group size ($b$=-0.13, $SE$=0.11, $p$=.22). Interpersonal distance did not differ across group familiarity ($b$=-0.023, $SE$=0.030, $p$=0.44) and group size ($b$=0.018, $SE$=0.021, $p$=.42). Participants with higher group familiarity had a lower number of users in FOV compared to those reporting lower group familiarity ($b$=-0.37, $SE$=0.15, $p$=.015). The number of users in FOV increased as group size increased ($b$=0.38, $SE$=0.11, $p$<.001). Finally, for the coverage of users in FOV, group familiarity ($b$=-0.059, $SE$=0.078, $p$=.45) and group size ($b$=0.040, $SE$=0.056, $p$=.48) were non-significant.

%For the models predicting the amount of virtual movement, there were non-significant effects for both group familiarity ($b$=0.29, $SE$=0.21, $p$=.18) and group size ($b$=-0.13, $SE$=0.11, $p$=.25). The average interpersonal distance did not differ across group familiarity ($b$=-0.37, $SE$=0.42, $p$=0.38) and group size ($b$=0.11, $SE$=0.22, $p$=.62). Furthermore, participants with high group familiarity exhibited a smaller number of users in FOV compared to those with low group familiarity ($b$=-0.59, $SE$=0.21, $p$=.0059). The number of users in FOV also increased as group size increased ($b$=0.37, $SE$=0.11, $p$=.0011). Finally, in predicting the coverage of other users in FOV, we found non-significant effects for group familiarity ($b$=-0.13, $SE$=0.11, $p$=.24) and group size ($b$=0.025, $SE$=0.058, $p$=.66).

\section{Discussion}
\subsection{Systematic Tradeoffs During Immersive Co-watching}

Across three experiments, we identified several tradeoffs, where viewing situations that yielded higher levels of engagement co-occurred with greater cognitive effort and workload.

\subsubsection{Spatial Boundary Type Shapes Social Presence and Mental Demand}

On \textbf{RQ1a}, Experiment 1 found that the spatial configuration affording the highest social presence toward co-watchers (i.e., box) was not the condition that yielded the lowest mental demand (i.e., reserved seating), implying that heightened social presence may entail increased cognitive demands. Different behavioral mechanisms likely informed these differences. For one, the patterns seen in FOV covered by user avatars across conditions matched those of the social presence of other users in Experiments 1 and 2, suggesting that users likely felt more socially present with others when other users took up more of their FOV. Interestingly, simply counting up the number of users in FOV did not reflect experiential differences, which we attribute to the nuances of overlapping avatars and interpersonal distance resulting in avatars with drastically different perceptual saliency. 

The differences in mental demand echoed the patterns seen in the amount of virtual movement: users who move more in the virtual environment tend to report greater mental demand. One interpretation is that participants who are 1) given less virtual space to move in and 2) placed in personal spatial boundaries rather than communal ones (i.e., reserved seating vs. box and general admission) did not need to engage in as much spatial and social negotiation and management, and therefore reported less mental demand. Given the exploratory nature of the present work confounding the two characteristics in the reserved seating configuration, it is unclear which mechanism primarily contributed to the experiential differences. 

Nevertheless, this work responds to Welsch et al.'s \cite{Welsch2021} call to study spatial negotiation in VR, with current works emphasizing how users actively negotiate personal space through their own behaviors \cite{Huang2022, Williamson2021}. Our findings add a complementary perspective: by shaping the spatial configurations of virtual environments such as the type of spatial boundaries and the amount of space made available to users, one can partially offload spatial and social negotiations from the individuals to the environment itself, thereby possibly modulating the mental effort required for social coordination. Importantly, as spatial boundaries and size of spatial constraints are confounded in Experiment 1, the findings here are to be read as differences between the spatial configurations as a whole, rather than claims about the effects of either confounding factor.

\subsubsection{Size of Allocated Personal Space as a Double-Edged Sword} \label{sec: disc exp 2}
Smaller allocated personal space and closer interpersonal distance were associated with higher social presence towards other users, and yielded greater disorientation and frustration (\textbf{RQ1b}). Several behavioral mechanisms could rationalize these differences. Similar to Experiment 1, higher social presence co-occurred with greater FOV coverage of user avatars, suggesting that when other users’ avatars take up greater perceptual space, users feel more socially present with others during co-watching. Several factors could contribute to the heightened frustration that accompanied the increase in social presence. To start, participants given less personal space lacked agency in managing the social and spatial context. They also experienced greater occlusion of the immersive story, as seen in their greater coverage of other users in their FOV, which may result in greater difficulty in viewing the stories. 

Another related factor concerns the diegetic misalignment between user avatars and the 360-degree story world. Because stories were rendered as skyboxes, the story world was spatially decoupled from the shared virtual environment occupied by users. Consequently, movements of other users in the shared virtual space were not perfectly integrated into the story world, and conversely, motion within the story world (e.g., motion of the story camera) did not reposition co-watchers in a way that preserved their relative spatial relationship with the story environment. This perceptual inconsistency may become more salient when users are positioned closer to one another, as nearby avatars that are poorly integrated with the story world can occupy greater portions of the visual FOV. Such heightened perceptual conflict may in turn exacerbate disorientation and frustration \cite{Busselle2008Model}, even as it increases awareness of and engagement with co-watchers. Our findings also align with research on breaks in presence, which can stem from internal inconsistencies and introduce greater workload \cite{Chandio2025, Putze2020}.

\subsubsection{Group Familiarity is Associated with Story Engagement and Workload}
To answer \textbf{RQ2}, Experiment 3 showed that participants reporting greater group familiarity also indicated greater attention towards the story and stronger spatial presence and social presence towards story characters. Because our manipulation check failed, these results reflect co-occurrence rather than directionality, though several explanations may help interpret these patterns. One is that individuals with high group familiarity tended to have fewer co-watchers in their FOV than those with low familiarity, which may signal lower effort in social monitoring, leaving more attentional resources available for the immersive story. This interpretation aligns with the equilibrium theory \cite{Argyle1965}, which posits that individuals regulate attention to maintain a desirable level of intimacy: individuals who are familiar with their social groups may require less visual monitoring to sustain social connection, whereas strangers may rely more on observing others. Consistent with this account, prior work indicates that greater group familiarity reduces the need to regulate social and task processes \cite{Janssen2009}, enabling users to allocate more attention to the primary task. 

That said, this positive relationship between group familiarity and attention towards the story was not mirrored in the one-item measure of attention towards others, as familiarity did not negatively predict it. We posit that this discrepancy can be due to measurement noise inherent to single-item questions, or it may indicate that attention is not zero-sum in our co-watching context: greater attention towards the story may not come at the expense of attending to co-watchers. 

Those who reported greater group familiarity also indicated higher perceptual and physical demand, which may reflect their greater attention towards immersive stories. Participants who paid more attention to the story and felt more immersed in the story world and socially engaged with its characters may have experienced greater workload to follow unfolding narrative events, scan the 360-degree visual environment, and actively maintain bodily orientation toward story elements. These findings extend the narrative engagement model \cite{Busselle2008Model}; greater self-reported group familiarity co-occurred with greater attentional focus on the immersive story and narrative transportation, as evidenced by the heightened spatial presence, social presence towards story characters, and workload as participants sustained mental models of the depicted narratives.

As familiarity was self-reported rather than manipulated, it is possible that our findings are explained by third variables. For example, participants who are more sociable and engaged may develop both stronger familiarity with others and attend more closely to the shared activities, producing the observed associations without either directly influencing the other.

\subsection{Practical Considerations and Outlooks for Co-watching of Immersive Stories}
\subsubsection{Crafting Spatial Configurations for Immersive Co-watching}

Our results highlight the importance of instantiating spatial boundaries in immersive co-watching environments to control for social engagement. Rather than treating space as a neutral or peripheral parameter, designers should leverage it as a design tool to regulate proxemics, guide viewing behavior, and balance social engagement with experiential demands.

When designing co-watching experiences, practitioners should not assume that larger spaces will uniformly increase virtual movement and interpersonal distance. In Experiment 2, providing participants with larger personal spaces to move in did not increase movement within those spaces. Instead, participants primarily used the additional space to position themselves farther from other users, resulting in viewing behaviors that resembled more traditional content-watching contexts (e.g., movie theaters, viewing immersive movies individually). In these settings, movement is either constrained (e.g., theaters) or perceptually unnecessary (e.g., position translation during 360-degree viewing does not alter the viewpoint). By contrast, the ability to move freely during immersive co-watching represents a novel affordance for many users. As such, practitioners should consider the intended role of virtual mobility during co-watching experiences and provide onboarding to help users understand when and how movement is meaningful. For instance, when the content is visually dense and warrants sustained attention, practitioners may consider implementing individual personal spaces (i.e., reserved seating) or large communal ones (i.e., general admission). Both configurations reduced the percentage of FOV occupied by other avatars compared to a small communal area (i.e., box), and individual personal spaces were additionally associated with the lowest mental demand of the three configurations.

Our results further suggest that boundary types systematically shape experiential outcomes. Configurations that emphasized personal boundaries were associated with lower mental demand, whereas communal boundaries can elicit stronger social presence toward co-watchers. Notably, Experiment 1 revealed a key distinction between virtual and physical spaces: while physical environments impose inherent spatial constraints, virtual environments can be extended indefinitely. Though moderately constrained communal spaces (e.g., 20×20 meter box) enhanced social presence, this benefit diminished when communal space was left largely unconstrained (e.g., 400×400 meter general admission). Therefore, for practitioners hoping to facilitate socially meaningful experiences, it may be beneficial to constrain the virtual space through spatial boundaries or implement features for summoning users \cite{Wood2023} or nudging interpersonal distance to a desirable range \cite{Wang2024}.

\subsubsection{Designing for Individual Differences in Social Context}
This work also highlights the importance of social context in shaping behavioral and psychological outcomes. Experiment 3 found that group familiarity shaped attention towards the story, spatial presence, social presence towards characters, and workload. Importantly, differences in group familiarity were quantified as an individual-level measure, meaning that the differences observed can occur within the same social group, where certain individuals felt closer to others, while others reported not knowing those in their group well at all. For practitioners, our results suggest that immersive co-watching systems should not be designed solely around group-level assumptions. Instead, individual users may experience the same social configuration in fundamentally different ways, with important consequences for attention allocation, presence, and cognitive and perceptual demands. Designs that implicitly assume high social comfort or familiarity among co-watchers may inadvertently increase workload for users who feel less socially connected, as these users may devote additional resources to monitoring, interpreting, or regulating social dynamics rather than engaging with the narrative content. Whenever possible, designers of social narrative experiences should support individual-level customization of spatial configurations, even for users within the same co-watching group. For example, platforms can implement features that allow individuals with varying group familiarity to selectively limit the number of users being rendered to reduce workload.

Our work also points to the potential of personalized and adaptive co-watching systems. Designers may consider features that automatically adjust the salience of social cues (e.g., avatar visibility, proximity) based on group familiarity with co-watchers and the cognitive demand of immersive stories \cite{Huang2021, Lindlbauer2019}. Personalized tutorials that account for differences in VR expertise, social norms, as well as familiarity with co-watchers could also lower workload. Similarly, systems could support adaptive spatial arrangements or viewing modes that accommodate preferences for social engagement, giving users control over how they balance narrative focus with social participation. When doing so, developers can apply transformed social interaction, which proposes that immersive systems can decouple tracked motions from rendered ones, such that all users can see a personalized version of the virtual scene \cite{Bailenson2005, Wang2024}. Finally, as individual- and group-level differences may alter over time, practitioners can consider building capabilities that support gradual transitions from lower to higher social engagement, or from individual to communal viewing modes.

\subsubsection{Beyond Watching 360-Degree Stories} While our work examined monoscopic 360-degree stories, we expect our findings to extend to other formats, though not uniformly. For example, regardless of content format, the presence of other avatars introduces an inevitable tension: our findings suggest that seeing and standing closer to co-watchers supports social presence, yet the same avatars occlude content and add visual clutter that can raise workload. This tradeoff is a structural feature of embodied co-presence rather than an artifact of 360-degree content, and we expect decisions about where co-watchers are positioned to matter in any format where content and companions compete for the same visual field. Even when co-watching 2D content, placement remains relevant, although attention is likely distributed differently when content occupies a more bounded region rather than surrounds the viewer.

Diegetic perceptual conflicts \cite{Gottsacker2021}, however, are likely to manifest differently across formats. In 2D screen-based viewing, the separation between user avatars and story content is clear. In stereoscopic 360-degree video, this separation depends on the congruency between the content's rendered IPD and that of the headset, and large mismatches may produce perceptual conflicts we did not observe in the monoscopic case. In volumetric and 3D cinematic content, where translation changes the viewpoint and users may be integrated into the story world, the patterns of occlusion and attention we observed may differ substantially. How much this matters is also content-dependent: content authored with co-watching in mind may anticipate where viewers gather in ways that solo-viewing content adapted for social contexts does not.

We position our implications as a starting point rather than a general account of co-watching. While we expect the tradeoff between social presence and content occlusion to generalize broadly, our findings remain tied to the medium we studied. Researchers and practitioners should therefore consider format closely before drawing on our implications.

\subsubsection{Co-Located vs. Remote Co-watching}

Our study was conducted with co-located groups, a choice driven largely by the practicalities of running large-group experiments within a classroom context, where coordinating up to ten students to follow study instructions remotely would have been considerably more difficult. Many social VR experiences, however, bring together users who are physically apart. We therefore consider here which of our findings are likely to transfer to remote settings and which may not. 

We expect findings based on the perceptual experience of co-watching to hold. For example, the association between social presence and visual occlusion will likely transfer to the remote context, as it is governed by what co-watchers see in VR rather than by their physical spatial arrangement. The same reasoning applies to the behavioral findings, as interpersonal distance, movement, and field-of-view coverage are properties of the virtual configuration rather than the physical one. Other findings warrant greater caution. Co-location shaped the social dynamics of our study in ways the virtual environment did not fully mediate, as participants could hear one another's voices and likely maintained an awareness of others that extended beyond the social cues their avatars provided. Remote co-watchers would depend far more heavily on avatar fidelity, nonverbal behavior, and spatialized audio to achieve comparable awareness. Social presence may be lower in remote settings, and perhaps be more sensitive to the virtual spatial configuration, as these channels would carry the full weight of social interaction. Seeing co-watchers within one's field of view may subsequently matter more, as avatars and their spatial audio become the only available social cues.

\section{Limitations and Future Works}
This work has several limitations. While our experiments varied spatial configurations and social context, they did not manipulate group size and composition. As these factors can shape social dynamics and attention \cite{Hammick2014, Wang2025}, research should examine their role in co-watching. Because Experiment 3 failed manipulation check, our group familiarity findings reflect associations rather than causal effects. Works could manipulate familiarity in other ways, such as comparing friend groups against strangers. Similarly, while the three spatial configurations in Experiment 1 were grounded in real-world use cases, boundary type and size of spatial constraint were confounded, making their individual effects difficult to isolate. Future work could cross these two dimensions to examine each individually and their potential interaction. Similarly, researchers can manipulate interpersonal distance and movement independently, isolating their individual effects that Experiment 2 was unable to separate. Avatar visibility and occlusion could likewise be varied on their own. Our research examined 360-degree content similar to \cite{Jun2020, Kim2022, Szita2018}, which impose specific perceptual and interaction constraints; researchers should investigate other forms of immersive media varying in interactivity and avatar embodiment \cite{Freeman2021, Latoschik2017}. 

Scholars should also examine populations beyond university students such as close friends and families, as well as strangers in public or semi-private virtual contexts. This is particularly important to contextualize our group familiarity findings, as familiarity was low in our sample (M=1.73 on a 5-point scale), with no participant reporting the highest level and only one reporting the second-highest, which was later winsorized. Our observed effects therefore reflect variation within a narrow, low-familiarity range, leaving open whether they hold at either extreme of social closeness. Measuring group familiarity on a single 5-point Likert scale also abstracts what distinguishes social relationships, including differing expectations and communication patterns among groups such as friends and colleagues. Groups that receive similar familiarity ratings may differ in how they co-watch, and characterizing these differences is an important direction for future work.

Our work is exploratory and focused on behaviors aggregated at the session level. Future studies should formulate theoretical predictions based on our findings to identify other mechanisms underlying engagement, motion sickness, and workload. Researchers can also examine moment-to-moment behaviors. While we observed little verbal communication during co-watching, research can examine verbal behaviors prior to and following co-watching, as well as more interactive stories that may prompt more verbal dialogues. Such approaches can be particularly valuable for designing co-watching agents for scaffolding attention and minimizing diegetic discrepancies \cite{Abe2021,Xu2019}, whose potential to shape social and psychological outcomes remains an open question. Finally, researchers should also consider outcomes such as narrative comprehension and memory recall, though both are stimulus-dependent and would require stories with comparable narrative structures. Spatial configuration preference and willingness to repeat the experience are likewise worth examining, and are perhaps better suited to a within-subjects design where participants can compare configurations directly. To this point, the field would further benefit from instruments developed and validated specifically for immersive co-watching.

\section{Conclusions}
Social VR increasingly enables people to watch immersive stories together, raising fundamental questions about how social and spatial factors reshape attention, engagement, and cognitive effort. The current work investigated how spatial configuration, size of allocated personal space, and group familiarity shaped immersive co-watching experiences. Across three experiments, we found that viewing scenarios that yielded greater social presence often elicited greater workload and motion sickness. For example, participants given a smaller allocated personal space reported greater social presence towards co-watchers, as well as heightened frustration and disorientation. Social context also mattered. Participants with greater group familiarity reported greater attentional focus, spatial presence and social presence towards story characters, and experienced greater perceptual and physical demand. The results of this work position immersive co-watching as a design space guided by tradeoffs rather than uniform experiential gains. By outlining systematic tradeoffs and design implications, we provide a foundation for future systems to support socially engaging yet cognitively manageable shared narrative experiences.

\section*{Acknowledgments}
We thank the ENGAGE team for their support throughout the study, as well as Robby Ratan and Gustav Bøg Petersen for their early feedback on study design and data analysis.
\bibliographystyle{IEEEtran}
\bibliography{references}
\section{Biography Section}
\vspace{-33pt}

\begin{IEEEbiography}[{\includegraphics[width=1in,height=1.25in,clip,keepaspectratio]{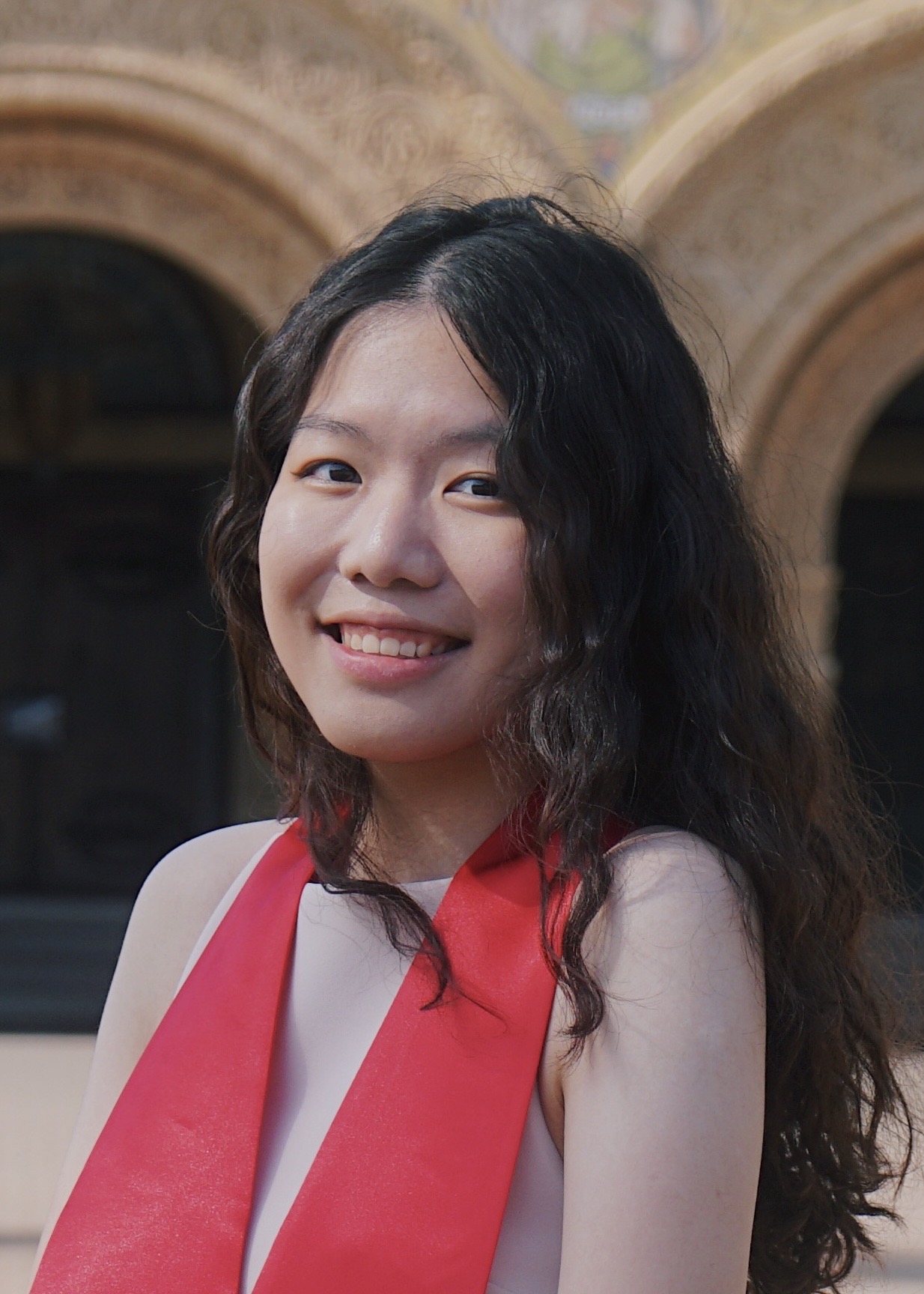}}]{Portia Wang} received the B.S. degree in Computer Science from Columbia University, New York, New York and her M.S. degree in Management Science and Engineering from Stanford University, Stanford, California. She is currently a Ph.D. Candidate in Communication at Stanford University, Stanford, California. Currently, she is investigating how individuals psychologically and behaviorally respond to immersive stories, and building tools that enable memory reliving using immersive technologies.
\end{IEEEbiography}
\vspace{-42pt}
\begin{IEEEbiography}[{\includegraphics[width=1in,height=1.25in,clip,keepaspectratio]{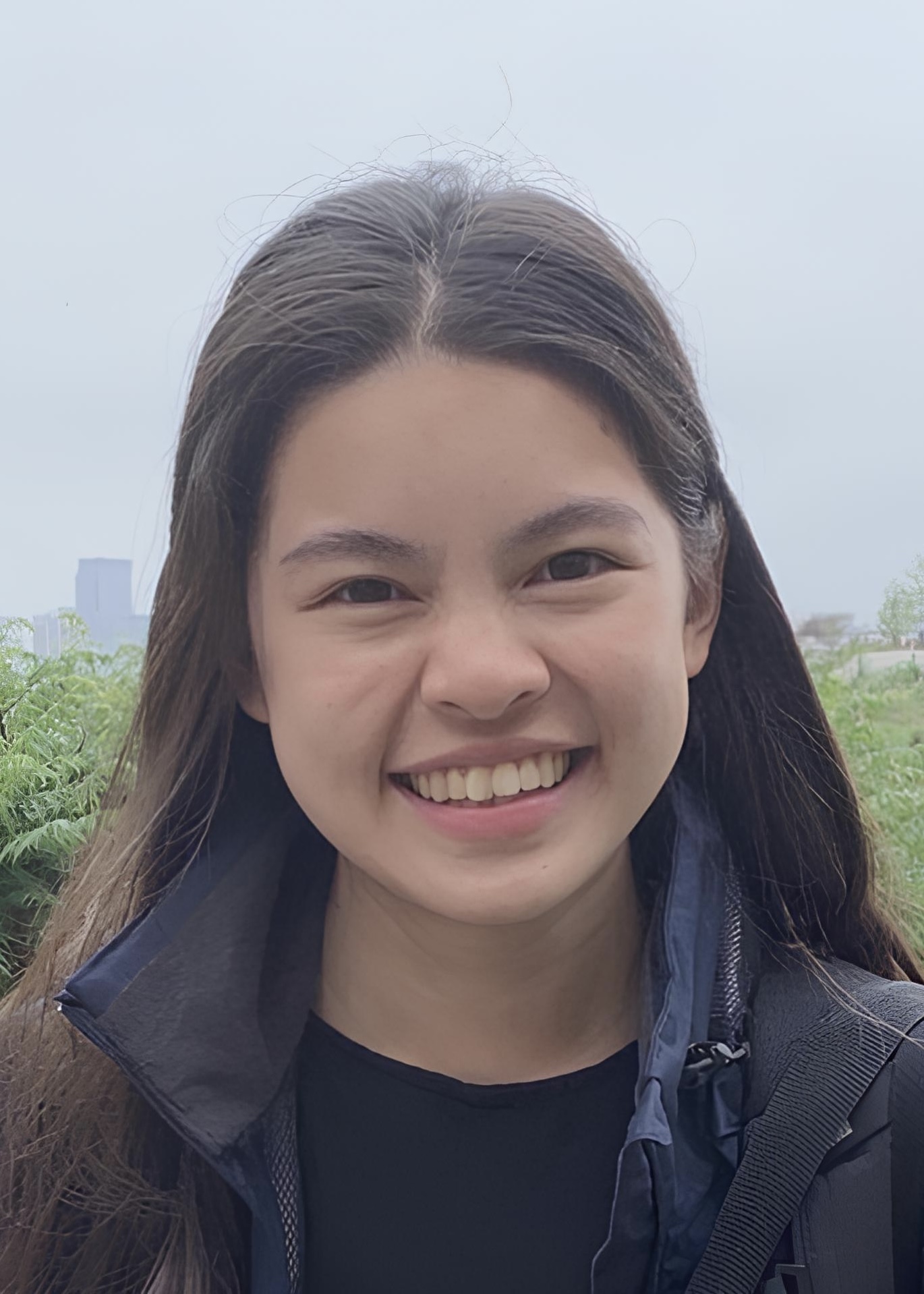}}]{Monique Santoso} received the B.A. degree in Environmental Science and Psychology from Middlebury College, Middlebury, Vermont and the M.A. degree in Communication from Stanford, Stanford, California. She is a Ph.D. Candidate in Communication at Stanford University. Her research focuses on the psychological, interpersonal, and behavioral implications of virtual reality, particularly as they relate to climate adaptation, prosocial attitudes and group interactions. 
\end{IEEEbiography}
\vspace{-42pt}
\begin{IEEEbiography}[{\includegraphics[width=1in,height=1.25in,clip,keepaspectratio]{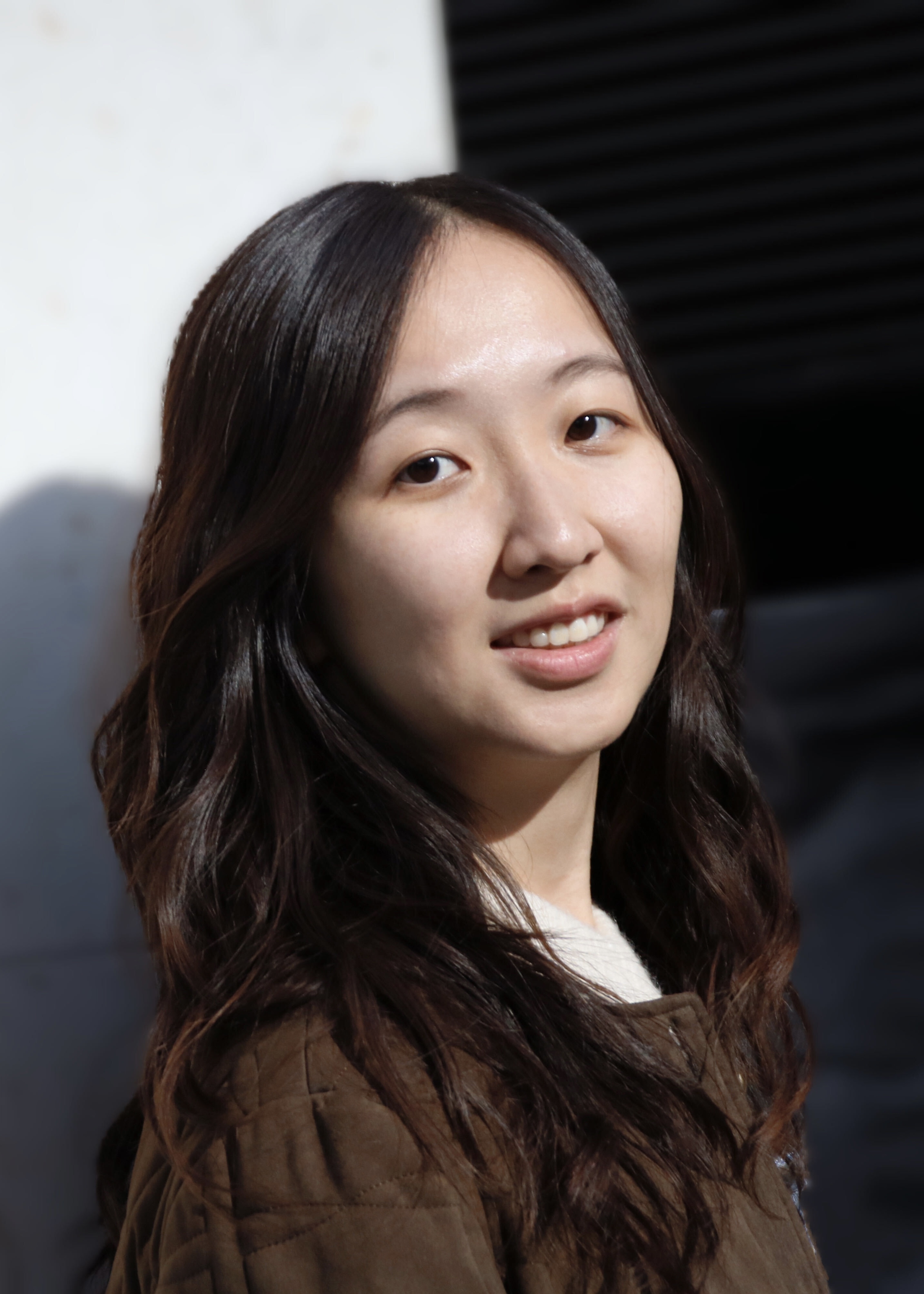}}]{Yujie Tao} recevied the B.A. in Journalism and Computer Science from the University of North Carolina at Chapel Hill, Chapel Hill, North Carolina and the M.S. degree in Computer Science from the University of Chicago, Chicago, Illinois. She is a Ph.D. Candidate in Computer Science at Stanford University, Stanford, California. Her research explores how computing systems reconnect users to everyday interactions by creating interfaces that augment our senses.
\end{IEEEbiography}
\vspace{-42pt}
\begin{IEEEbiography}[{\includegraphics[width=1in,height=1.25in,clip,keepaspectratio]{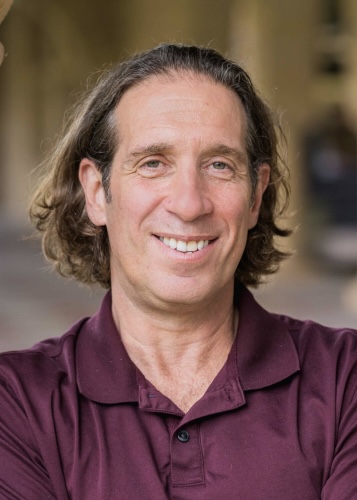}}]{Jeremy Bailenson} received the B.A. degree from the University of Michigan, Ann Arbor, Michigan, and the Ph.D. degree in cognitive psychology from Northwestern University, Evanston, Illinois. He is founding director of Stanford University’s Virtual Human Interaction Lab, Thomas More Storke professor with the Department of Communication, professor (by courtesy) of Education, professor (by courtesy) program in Symbolic Systems, a senior fellow at the Woods Institute for the environment, and a faculty leader at Stanford’s Center for Longevity. %He spent four years at the University of California, Santa Barbara, Santa Barbara, California as a postdoctoral fellow and then an assistant research professor.
\end{IEEEbiography}
%TC:endignore

\vspace{11pt}

\vfill

\end{document}